\documentclass[twocolumn,aps,prl]{revtex4-1}

	\usepackage[usenames,dvipsnames]{xcolor}
	\usepackage{amsmath}
	\usepackage{amsfonts}
	\usepackage{amssymb}
	
	\usepackage[colorlinks=true,citecolor=blue,linkcolor=red]{hyperref}
	\usepackage{graphicx}
	\usepackage{bbold}					
	\usepackage[makeroom]{cancel}		
	\usepackage{multirow}				
	\usepackage[normalem]{ulem}        
	\usepackage{array}

	\newcolumntype{x}[1]{>{\centering\let\newline\\\arraybackslash\hspace{0pt}}p{#1}}

	\DeclareMathAlphabet{\mathbbold}{U}{bbold}{m}{n}

	\newcounter{subeqn} %
	\makeatletter
	\@addtoreset{subeqn}{equation}
	\makeatother

\definecolor{TB}{rgb}{1,0.5,0}

\definecolor{ZX}{rgb}{0.4,0,1}
\definecolor{ZX}{rgb}{0,0,0}
\def\ZX#1{{\color{ZX}#1}}

\definecolor{HIGHL}{rgb}{1.0, 0.0, 0.5}
\definecolor{HIGHL}{rgb}{0.0, 0.0, 0.0}

\begin{document}
\title{Quantum entanglement and modulation enhancement of free-electron--bound-electron interaction}

\date{\today}

\begin{abstract}

The modulation and engineering of the free-electron wave function bring new ingredient to the electron-matter interaction. We consider the dynamics of a free-electron passing by a two-level system fully quantum mechanically and study the enhancement of interaction from the modulation of the free-electron wave function. 
In the presence of resonant modulation of the free-electron wave function, we show that the electron energy loss/gain spectrum is greatly enhanced for a coherent initial state of the two-level system. Thus, a modulated electron can function as a probe of the atomic coherence. We further find that distantly separated two-level atoms can be entangled through interacting with the same free electron. 
Effects of modulation-induced enhancement can also be observed using a dilute beam of modulated electrons. 
\end{abstract}
\author{Zhexin Zhao$^1$}
\author{Xiao-Qi Sun$^{2,3}$}
\email[]{xiaoqi20@illinois.edu}
\author{Shanhui Fan$^1$}\email[]{shanhui@stanford.edu}
\affiliation{$^1$Department of Electrical Engineering, Ginzton Laboratory, Stanford University, Stanford, CA 94305, USA}
\affiliation{$^2$Department of Physics, McCullough Building, Stanford University, Stanford, CA 94305, USA}
\affiliation{$^3$Department of Physics, Institute for Condensed Matter Theory,
University of Illinois at Urbana-Champaign, IL 61801, USA}
\maketitle

\emph{Introduction.--} 
The interactions of free electrons with matters have provided a number of technologies for the study of material and photonic systems \cite{zewail2010four, de2010optical, polman2019electron, di2019probing}. 
Among these technologies, electron energy loss spectroscopy (EELS) probes the excitation spectrum of the matter \cite{zewail2010four, de2010optical, polman2019electron, nazarov1999multipole, nazarov2016role, nazarov2017probing}, and photo-induced near-field electron microscopy (PINEM) detects the near-field of nano-structures under optical pumping \cite{barwick2009photon, garcia2010multiphoton, park2010photon, feist2015quantum, priebe2017attosecond, dahan2019observation, reinhardt2020theory, rivera2020light}.
Moreover, in recent years, the quantum nature of the free electron has attracted considerable research interest \cite{polman2019electron, di2019probing, dahan2019observation, barwick2009photon, garcia2010multiphoton, park2010photon, feist2015quantum, priebe2017attosecond, dahan2019observation, reinhardt2020theory, rivera2020light, vanacore2018attosecond, vanacore2019ultrafast, echternkamp2016ramsey, pan2019spontaneous, remez2019observing, larocque2018twisting, madan2020quantum, zhou2019quantum}. In particular, over the last decade, engineering the quantum states of the free electron becomes possible. For example, experiments demonstrated the modulation of a single electron wave function by ultra-fast laser technique \cite{barwick2009photon, garcia2010multiphoton, park2010photon, feist2015quantum, priebe2017attosecond, dahan2019observation, reinhardt2020theory, rivera2020light, vanacore2018attosecond, vanacore2019ultrafast, echternkamp2016ramsey}. 

With the capability of wave function engineering, it is crucial to investigate how such quantum engineering can be used to enhance and tailor electron-material interaction and to create new functionalities.
Studies have shown increased scattering cross-section, and coherent control of atomic transitions using a resonant modulated electron beam \cite{favro1971energy, gover2020free, ratzel2020quantum}. 
Entanglement between electron and photons has been studied in the interaction of an electron with an optical cavity \cite{kfir2019entanglements}. Moreover, recently Gover and Yariv has studied the interaction of a modulated electron beam and an atom, in a semiclassical formalism, and showed the possibilities of Rabi oscillation in such free-electron-bound-electron interaction \cite{gover2020free}. 

In this letter, we apply a fully quantum scattering matrix description to the free-electron--bound-electron interaction \cite{de2010optical, polman2019electron, rivera2020light, ritchie1988inelastic, ratzel2020quantum}.
We investigate the electron energy loss/gain spectrum and the perturbation on the density matrix of the two-level system, highlighting the enhancement of the interaction resulted from the modulation of the free electron. We find the potential of probing the coherence of the two-level system using the modulated free electron.
The quantum treatment also predicts possibility of using multiple scattering processes to generate entanglement between two bound-state electrons. Finally, we consider the interaction between the two-level system and a dilute beam of modulated electrons. We discuss a possible partially mixed steady state of the atom due to the modulation, while also provide rigorous foundation for the semi-classical results of the Rabi oscillation of the two-level system~\cite{gover2020free}.

\begin{figure}
    \centering
    \includegraphics[width=0.9\linewidth]{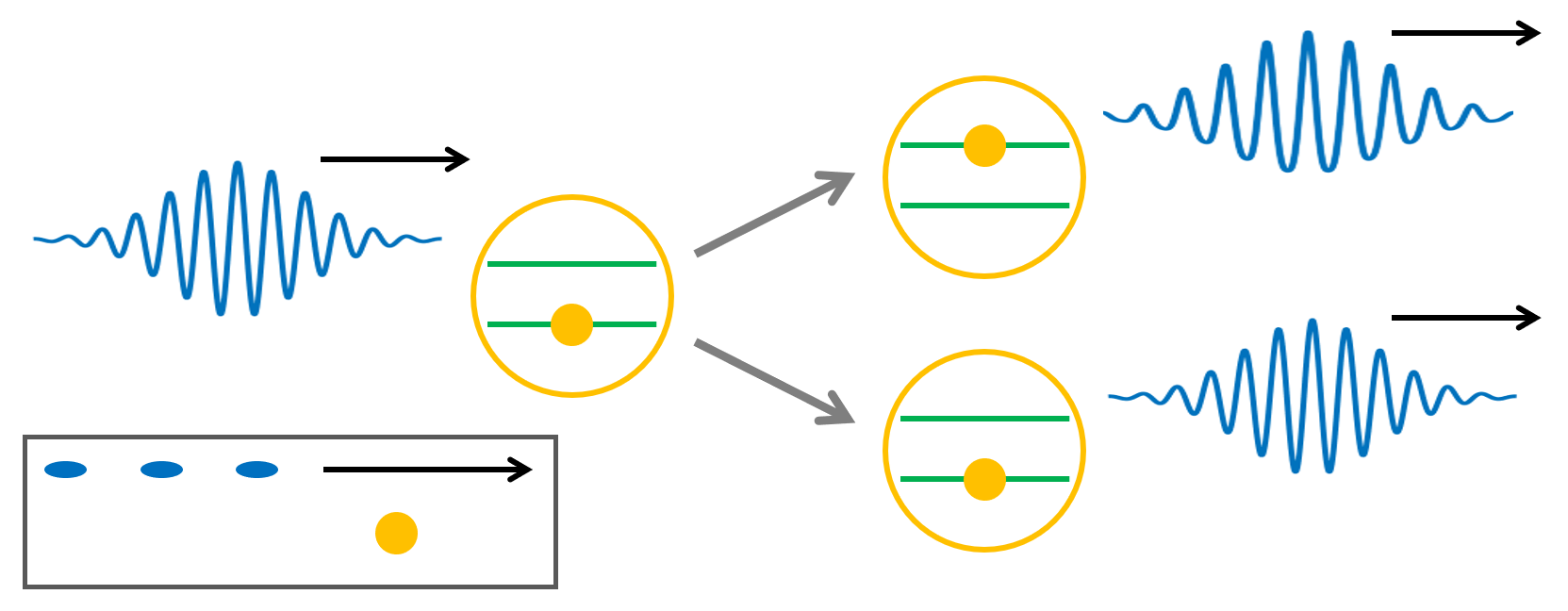}
    \caption{Schematic of the interaction between an electron (blue wave packet) and a two-level system (orange circle). The insert shows a schematic of a train of electrons interacting with the two-level atom.}
    \label{fig:schematic_electron_atom}
\end{figure}

\emph{Model setup.--} We consider the scattering problem between a free electron with a two-level system from first principle (Fig. \ref{fig:schematic_electron_atom}). In the low-velocity limit, the electrons are governed by Schrodinger equation and the interaction potential is the Coulomb potential. 
In analyzing the Coulomb interaction, we consider only the interaction between the free electron and the bound electron, and use dipole approximation.
If we further assume that the transverse distribution of the free-electron wave function is unchanged, the model Hamiltonian becomes:
\begin{equation}
    \label{eq:Hamiltonian_e-a}
    \begin{split}
    H = &\; \sum_\alpha E_\alpha c_\alpha^\dagger c_\alpha + \sum_{k}E^\textrm{free}_k c_k^\dagger c_k \\ & + \sum_{q} b_{q}[g_{21}(q)\sigma^+ + g_{12}(q)\sigma^-],
    \end{split}
\end{equation}
where $\alpha = 1,2$ represents the bound states of the two-level system, $c^\dagger$ and $c$ are the creation and annihilation operators, $\sigma^+=c_2^\dagger c_1$, $\sigma^-=c_1^\dagger c_2$, and $b$ is the electron ladder operator $b_q=\sum_{k}c_{k-q}^{\dagger}c_{k}$ \cite{feist2015quantum, kfir2019entanglements}. 
The coupling between the free electron and the bound state electron ($g_{ij}$) can be generally derived from the Coulomb interaction (Supplemental Material (SM) Sec.\ I \cite{supp_qebeam}).

We treat the free-electron--bound-electron scattering problem perturbatively, since the interaction is generically weak. Under the weak coupling assumption, only the electron ladder operator with $q = \omega_a/v_0$ matches the two-level system transition and contributes to the scattering. Here, $\omega_a$ is the transition frequency of the two-level system and $v_0$ is the velocity of the electron wave packet. In the following, we consider only such matched electron ladder operator and omit the subscript $q$. To second order in the dimensionless coupling coefficient $g$, which is proportional to $g_{21}(\omega_a/v_0)$, the scattering matrix is \cite{supp_qebeam}
\begin{equation}
    \label{eq:scattering_matrix}
    S \approx \big(1-\frac{1}{2}|g|^2\big)I -i(gb\sigma^+ + g^*b^\dagger \sigma^-),
\end{equation}
where $I$ is the identity operator.
The magnitude of the coupling ($g$) depends on the transition dipole moment, the free electron velocity and transverse distance. A typical value is $|g|\sim 10^{-3}$. An estimation for the Tin-Vacancy (SnV) \cite{trusheim2020transform, rugar2020narrow} is in SM Sec.\ II \cite{supp_qebeam}.
This concise scattering matrix (Eq. (\ref{eq:scattering_matrix})) provides a fully quantum mechanical description of the interaction between the free electron and the two-level system. It manifests the entanglement between the atomic transition of the bound electron and the hopping on the energy ladder of the free electron. It works well for arbitrary initial states and reveals results due to free-electron modulation that are obscured in previous studies.

\begin{figure}
    \centering
    \includegraphics[width=0.99\linewidth]{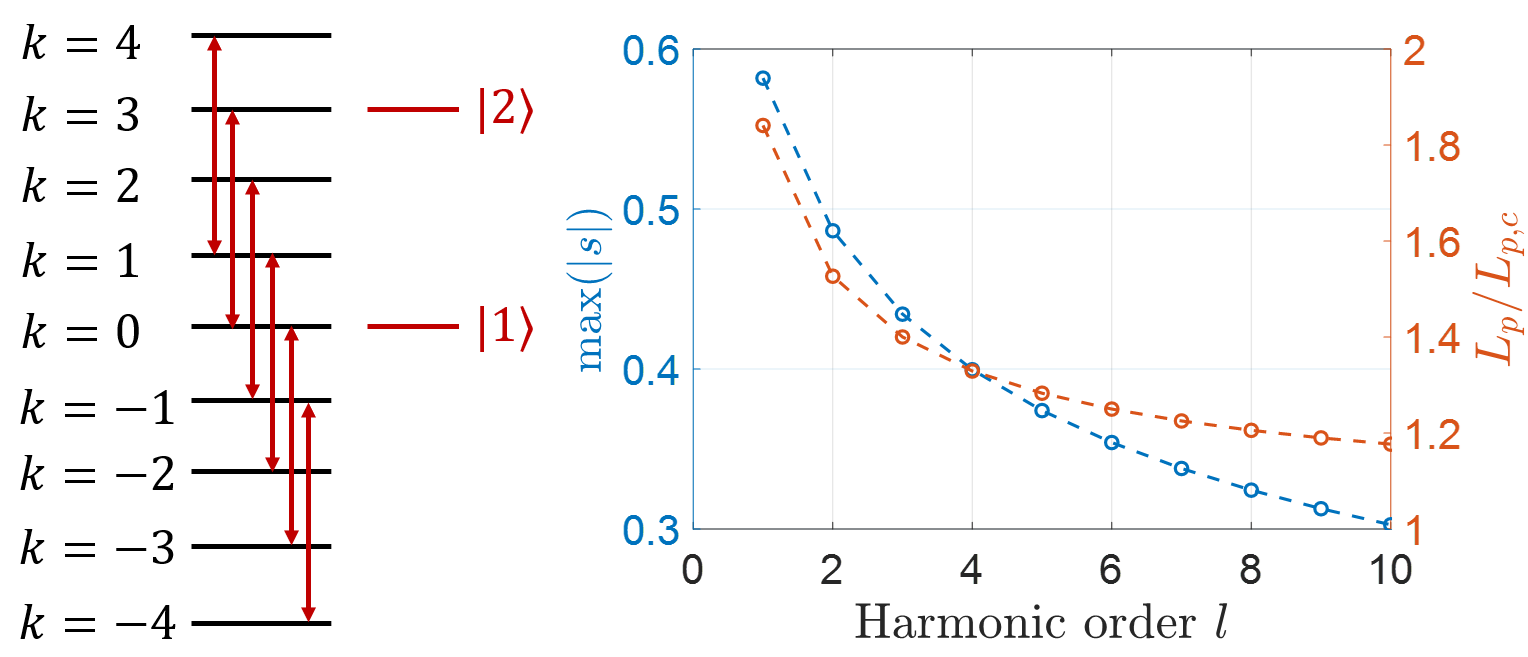}
    \caption{Left: Schematic of the electron ladder for an electron modulated at frequency $\omega$ and the energy levels of the two-level system, where $\omega_a = l\omega$ and $l=3$. Right: Maximal $|s|$ and the corresponding drift length for different harmonic orders.}
    \label{fig:modulation_harmonics}
\end{figure}

\emph{Modulation of the free electron.--} We briefly revisit the free-electron energy modulation and wave function engineering \cite{feist2015quantum}. In PINEM, the free electron interacts with the near-field and absorbs or emits integer number of photons. The free-electron wave function with small initial energy spread then contains multiple energy components. The multiple energy levels are on a ladder with energy separation equal to the absorbed/emitted photon energy $\hbar \omega$ (Fig. \ref{fig:modulation_harmonics}). The unitless modulation strength $g_m$ is defined as $g_m = e\int dz \exp(-i\omega z/v_0)E_{mz}(z)/2\hbar\omega$, where $z$ is the electron propagation direction and $E_{mz}$ is the $z$-component of the modulation field. In the free-drift region after the energy modulation, different energy components accumulate different additional phases and the real-space electron probability distribution forms a train of micro-bunches \cite{feist2015quantum}. Such free electron with engineered wave function is referred to as the modulated free electron in this study.

\emph{Perturbation on the two-level system.--}
\ZX{The modulated free electron can drive the transition of the two-level system coherently.} 
The change in the density matrix of the two-level system is obtained by tracing out the free electron state, \ZX{$\Delta \rho_a = \textrm{Tr}_e(S\rho_{ea}^i S^\dagger - \rho_{ea}^i)$}, 
where the initial density matrix $\rho_{ea}^i = \rho_{e}^i\otimes\rho_a^i$. In terms of the two-level system density matrix elements, we get
\begin{equation}
    \label{eq:density_matrix_elements}
    \begin{split}
    &\Delta \begin{pmatrix} \rho_\textrm{a11} \\ \rho_\textrm{a22} \\ \rho_\textrm{a12} \\ \rho_\textrm{a21} \end{pmatrix} = -iM \begin{pmatrix} \rho_\textrm{a11} \\ \rho_\textrm{a22} \\ \rho_\textrm{a12} \\ \rho_\textrm{a21} \end{pmatrix}, \\
    & M = \begin{pmatrix} -i|g|^2 & i|g|^2 & -gs & g^*s^* \\ i|g|^2 & -i|g|^2 & gs & -g^*s^* \\ -g^*s^* & g^*s^* & -i|g|^2 & ig^{*2}s^*_2 \\ gs & -gs & ig^2s_2 & -i|g|^2 \end{pmatrix},
    \end{split}
\end{equation}
where $s = \langle b \rangle$ and $s_2 = \langle b^2 \rangle$ are parameters determined by the state of the incident electron.

\ZX{The lower-left off-diagonal $2\times 2$ block of $M$ represents the change in the atomic coherence to the first order of $g$, which depends strongly on the incident electron state. For a quasi-monochromatic electron or a modulated electron without drift for micro-bunching, $s=0$.}
However, for a modulated electron with a proper drift length for micro-bunching \cite{feist2015quantum, black2019net, schonenberger2019generation, morimoto2018diffraction}, $s\neq 0$ and may approach unity with sophisticated modulation \cite{priebe2017attosecond, reinhardt2020theory}. \ZX{Therefore, the induced coherence in the two-level atom is controlled by the modulation on the free electron.}

\ZX{We further study $s$ for a modulated electron and discuss conditions to maximize $|s|$ in typical PINEM.}
We consider a Gaussian electron wave packet with momentum spread $\sigma_q \ll \omega_a/v_0$. The wave packet is modulated by a laser at frequency $\omega$, with modulation strength $g_m$, and propagates for a drift length $L_p$ before it interacts with the two-level system (SM Sec.\ III \cite{supp_qebeam}). We find that the magnitude of $s$ is maximized on resonance $\omega_a = l\omega$, where the harmonic order $l$ is an integer. Under this resonance condition, 
\begin{equation}
    \label{eq:s_modulation}
    |s|\approx \Big| J_l\Big[4|g_\textrm{m}|\sin\Big(\frac{l L_p}{4|g_\textrm{m}|L_{p,c}}\Big)\Big]\Big|,
\end{equation}
where $L_{p,c}$ is the drift length for perfect bunching from classical analysis (SM Sec.\ III \cite{supp_qebeam}). The maximal $|s|$ is the peak of Bessel function $J_l$ and the argument gives the optimal drift length. \ZX{To reach the maximal $|s|$ for the $l$th harmonic, the modulation strength $|g_m|>l/4$.} When the modulation is strong, 
$|s|\approx J_l(lL_p/L_{p,c})$, \ZX{the optimal drift length becomes independent of $g_m$.} The maximal $|s|$ and the corresponding optimal drift length are shown in Fig. \ref{fig:modulation_harmonics}. We find that $|s|$ decreases slowly with increasing harmonic order. This trend indicates the potential to drive two-level system at high harmonics with the modulated electron \cite{gover2020free}. The optimal drift length is larger than the drift length to create the shortest electron bunch (Fig. \ref{fig:modulation_harmonics}), especially for small harmonic orders.

\emph{Electron energy loss/gain spectrum.--} 
\ZX{The atomic coherence can be probed by the modulated free electron in EELS, with an enhanced signal.}
The change in the density matrix of the free electron is
\begin{align}
    \label{eq:delta_rho_e}
    \begin{split}
    \Delta\rho_e = &\; \textrm{Tr}_a(S\rho^i_{ea}S^\dagger - \rho^i_{ea}) \\ = &\; -|g|^2\rho^i_e + |g|^2\rho^i_{a11} b \rho^i_e b^\dagger + |g|^2\rho^i_{a22}b^\dagger \rho^i_e b \\ & -i[g\rho^i_{a12}(b \rho^i_e - \rho^i_e b) + g^*\rho^i_{a21}(b^\dagger \rho^i_e - \rho^i_e b^\dagger)].
    \end{split}
\end{align}
Thus, the free-electron spectrum change is $\Delta \rho_e(k) = \langle k|\Delta \rho_e|k \rangle$ (SM Sec.\ III \cite{supp_qebeam}). The average energy change is
\begin{equation}
    \label{eq:electron_average_energy_change}
    \begin{split}
    \langle \Delta E_e\rangle = &\; \hbar \omega_a |g|^2(\rho^i_{a22} - \rho^i_{a11})\\ & +i\hbar\omega_a(g\rho^i_{a12}\langle b \rangle - g^* \rho^i_{a21}\langle b^\dagger \rangle).
    \end{split}
\end{equation}
We find that the energy exchange between the free electron and the two-level system depends on the initial states of the two-level system and the free electron. 
When $\langle b\rangle \neq 0$, the electron energy loss, as well as the energy spectrum variance, is proportional to the off-diagonal elements of the two-level system density matrix to the first order in $g$. 
On the contrary, for an electron in a totally mixed state or being quasi-monochromatic, 
\ZX{$\Delta\rho_e(k)$ and $\langle \Delta E_e\rangle$ become proportional to $|g|^2$.}
Thus, the control of the free-electron state is crucial to observe the first order effects. In short, Eqs. (\ref{eq:delta_rho_e}) and (\ref{eq:electron_average_energy_change}) imply an opportunity to probe the coherence of the two-level system by measuring the energy loss/gain spectrum of the free electron.

\begin{figure}
    \centering
    \includegraphics[width=0.84\linewidth]{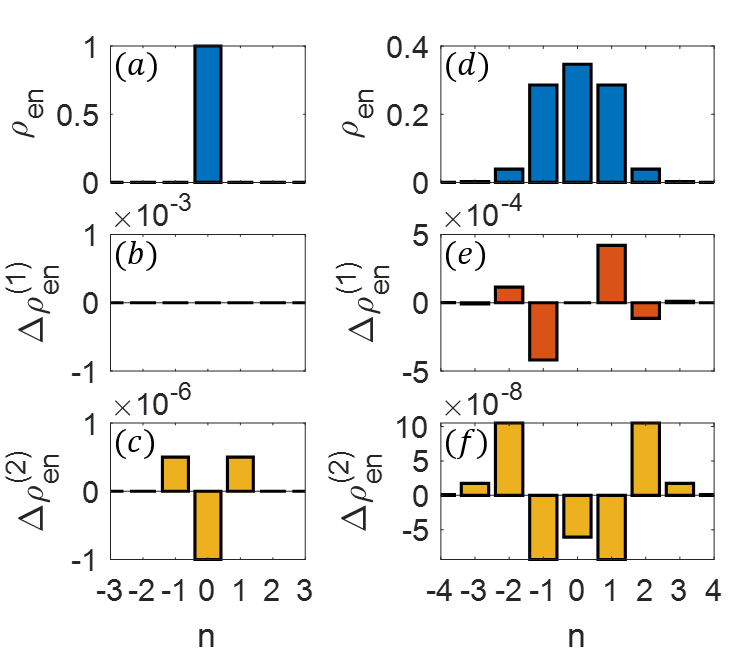}
    \caption{Spectrum change of the free electron interacting with a two-level system in state $|\Psi_a\rangle = (|0\rangle + |1\rangle)/\sqrt{2}$, where the free electron is either a 60 keV Gaussian wave packet with $\sigma_p \ll \omega_a/v_0$ (a-c) or a resonant modulated wave packet with optimal modulation strength $g_m=0.68$, modulation frequency $\omega_a$ (620 nm) and drift length 10 mm (d-f). (a), (d) show the electron energy spectrum before the interaction. (b), (e) and (c), (f) are the spectrum changes after the interaction in the first and second order of $|g|$, respectively, where a typical $g = 1\times 10^{-3}$ is assumed.}
    \label{fig:EELS}
\end{figure}

The enhanced EELS signal that is to the first order of $g$ can be optimized by controlling the free-electron wave function. We discuss the optimal conditions for a typical PINEM modulated electron in SM Sec.\ IV \cite{supp_qebeam}. As an demonstration, we study the free-electron spectrum change of a modulated free electron interacting with a two-level system (SnV) in a superposition state $|\Psi_a\rangle = (|0\rangle + |1\rangle)/\sqrt{2}$ (Fig. \ref{fig:EELS}). We assume that the 60 keV electron is modulated at resonant frequency $\omega_a$ (620 nm) with optimal strength $g_m = 0.68$ and drift length $L_p = 10$ mm \cite{supp_qebeam}. Before interacting with the two-level system, the electron spectrum is shown in Fig. \ref{fig:EELS}(d), which is a typical PINEM spectrum \cite{feist2015quantum}. After the interaction, the spectrum change that is proportional to $|g|$ ($|g|^2$) is plotted in Fig. \ref{fig:EELS}(e) (Fig. \ref{fig:EELS}(f)), where we assume a typical value $g = 1\times 10^{-3}$. In this case, the $|g|^2$ contribution is much smaller than the $|g|$ contribution and can be neglected. The spectrum change is as Fig. \ref{fig:EELS}(e). However, if either the free electron or the two-level system loses the coherence, the $|g|$ contribution disappears and the spectrum change is as Fig. \ref{fig:EELS}(f). In comparison, we study a Gaussian wave packet with small initial energy spread ($\sigma_p \ll \omega_a/v_0$) (Fig. \ref{fig:EELS}(a)) interacting with the same two-level system in state $|\Psi_a\rangle$. After the interaction, the spectrum change is proportional to $|g|^2$, with zero contribution proportional to $|g|$ (Fig. \ref{fig:EELS}(b-c)). 
In contrast to typical PINEM spectra, the spectrum change probing the atomic coherence can be anti-symmetric (Fig. \ref{fig:EELS}(e)). This anti-symmetric part can be extracted to determine the atomic coherence experimentally (SM Sec.\ V \cite{supp_qebeam}).

When the two-level system is initially at the ground state and the incident free electron has a well-defined momentum $\hbar k_0$, our result is consistent with conventional EELS \cite{de2010optical}: $\Delta \rho_e (k) = -|g|^2 \delta_{k,k_0} + |g|^2 \delta_{k, k_0-\omega_a/v_0}$, which is proportional to $|g|^2$. Thus, in typical situations $|g|\ll 1$, the modulated electron provides an opportunity to probe the coherence of the two-level system with an EELS signal much stronger than the conventional EELS signal.

\begin{figure}[t]
    \centering
    \includegraphics[width=0.8\linewidth]{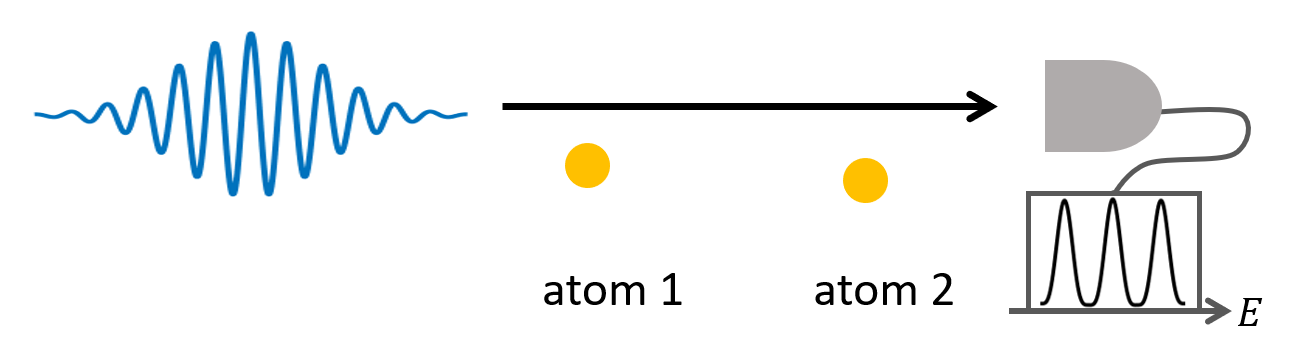}
    \caption{Schematic of an electron interacting with two atoms.}
    \label{fig:schematic_two_atom_entanglement}
\end{figure}

\emph{Entanglement.--} Two separated two-level atoms can be entangled through the interaction with the same free electron. As illustrated in Fig. \ref{fig:schematic_two_atom_entanglement}, a single electron interacts in sequence with two-level atoms 1 and 2, which have the same transition frequency $\omega_a$. After the interactions, the electron energy is measured. We assume that the two atoms are separated enough to avoid electron wave function overlap between the two atoms.
Suppose that both atoms are at the ground state and the free electron has energy spread $\sigma_q\ll\hbar\omega_a$ before the interaction, and the coupling coefficient between the electron and atom 1 (2) is $g_1$ ($g_2$). The final state after scattering is
\begin{equation}
    |\Psi_{e12}^f\rangle = [S_{2}(g_2)\otimes I_{1}][S_{1}(g_1)\otimes I_{2}]|\Psi_e^i\rangle |\Psi_1^i\rangle |\Psi_2^i\rangle ,
\end{equation}
where $S_1$ ($S_2$) acts on the product space of the electron and atom 1 (2). We assume that when the electron interacts with one of them, the other atom has no influence. An explicit form of the final state is shown in SM Sec.\ VI \cite{supp_qebeam}.

After the interactions, if certain energy of the electron is measured, the two atoms are in an entangled state. For instance, when the electron has initial momentum $\hbar k$ and final momentum $\hbar k - \hbar\omega_a/v_0$, the two atoms are in the entangled state
\begin{equation}
    \label{eq:entangled state}
    \frac{1}{\sqrt{|g_1|^2 + |g_2|^2}}\Big[g_2|\Psi_{1}^i\rangle \left(\sigma^+|\Psi_{2}^i\rangle\right) + g_1 \left(\sigma^+|\Psi_{1}^i\rangle\right) |\Psi_{2}^i \rangle\Big].
\end{equation}
The probability of obtaining such entangled state is $\sim (|g_1|^2+|g_2|^2)$. It is possible to increase the coupling coefficients and hence the probability of entangled state generation by \ZX{choosing two-level systems with large transition dipole moment,} decreasing the distance between the electron trajectory and atoms, and decreasing the electron speed \cite{cox2020nonlinear}. Furthermore, entanglement between multiple atoms is possible if the single electron interacts with each of them before the energy measurement.

\emph{Multiple electrons.--}
The enhancement of perturbation on a two-level system resulting from the free-electron resonant modulation is also manifested in the interaction between the two-level system and a dilute electron beam (inset of Fig. \ref{fig:schematic_electron_atom}), \ZX{since the atomic coherent excitation can build up in the sequential scattering of the electrons by the atom.}
The adjacent electrons are separate with an average separation larger than the longitudinal size of the electron wave function, such that, for each electron interacting with the two-level system, the influence of other electrons is negligible. 
Since the perturbation of a single free-electron interaction is given by Eq. (\ref{eq:density_matrix_elements}), the effective dynamics of the two-level system is
\begin{equation}
    \label{eq:atom_effective_dynamics}
    \frac{\Delta u}{\Delta t} = -i\frac{M}{T}u - \Gamma u, \; \Gamma= \begin{pmatrix} 0 & -\frac{1}{\tau} & 0 & 0 \\ 0 & \frac{1}{\tau} & 0 & 0 \\ 0 & 0 & \frac{1}{2\tau} & 0 \\ 0 & 0 & 0 & \frac{1}{2\tau} \end{pmatrix},
\end{equation}
where $u = [\rho_{a11},\, \rho_{a22},\, \rho_{a12},\, \rho_{a21}]^T$, $T$ is the time separation between adjacent electrons, and $\tau$ is the decay time of the two-level system \cite{laussy2009luminescence}. The average dynamics is similar to the optical Bloch equations \cite{ratzel2020quantum}. The effective Hamiltonian describing the two-level system driven by the electron beam is $H_\textrm{eff} = M/T-i\Gamma$. 
The effective Hamiltonian has a zero eigenvalue corresponding to the steady state: 
\begin{equation}
    \label{eq:steady_state}
    \rho_a = \frac{1}{1+2\big(\frac{2|gs|\tau}{T}\big)^2}\begin{bmatrix} 1+\Big(\frac{2|gs|\tau}{T}\Big)^2 & i\frac{2g^*s^*\tau}{T} \\ -i\frac{2gs\tau}{T} & \Big(\frac{2|gs|\tau}{T}\Big)^2 \end{bmatrix},
\end{equation}
where we assume $|s| \gg |g|$ due to resonant modulation. Without modulation, i.e. $s = s_2= 0$, the steady state becomes $\rho_a = \frac{1}{1+2|g|^2\tau/T} \begin{bmatrix} 1+|g|^2\tau/T & 0 \\ 0 & |g|^2\tau/T \end{bmatrix}$, where the off-diagonal elements are zero and the excited state probability is typically much smaller in comparison with the modulated case. Thus, the analysis of the two-level system steady state shows the enhanced interaction due to the modulation of the free electron.

Before reaching the steady state, the two-level system may experience Rabi oscillation if $T<8|gs|\tau$, which can be observed with state-of-the-art experimental technology (SM Sec.\ VII \cite{supp_qebeam}). In the limit of small two-level-system decay rate, the Rabi oscillation frequency is $\Omega_R = 2|gs|/T$, which is consistent with the semi-classical results \cite{gover2020free}. The physical intuition about this consistency is that the classical interpretation of $s = \langle b\rangle$ is the amplitude of the current distribution with spatial frequency $k = \omega_a/v_0$ \cite{rivera2020light}. Therefore, our theory provides a rigorous foundation for the semi-classical results.

\emph{Conclusion.--} We present a quantum description for the interaction between a free electron and a two-level system that reveals the quantum entanglement and applies to the modulated electron straightforwardly. 
We highlight the enhancement of interaction due to the modulation of the free-electron wave function\ZX{, which can be utilized to probe the atomic coherence}. Such enhancement persists when the two-level system interacts with a dilute modulated electron beam. We also discuss an approach to create entanglement between distant two-level systems through the interaction with the same free electron. 
Our study of the free-electron--bound-electron interaction emphasizes the significance of free-electron wave function engineering and provides new perspectives for ultra-fast physics studies using the free-electron probe.

\ZX{Immediately after the submission of our manuscript, several papers \cite{ruimy2020towards, gover2020resonant, wong2020control, de2020quantum} appeared on the preprint archive, providing quantum treatments of interactions of modulated electrons with the atom.}

\begin{acknowledgments}
\emph{Acknowledgments.--} We thank Prof.\ Meir Orenstein, Mr.\ Dylan Black, Dr.\ Jean-Philippe MacLean, Ms.\ Alison Rugar, Dr.\ Rahul Trivedi, Fan group members and members of the ACHIP collaboration for helpful discussions and suggestions. This work is supported by Gordon and Betty Moore Foundation (GBMF4744). Z.\ Zhao acknowledges Stanford Graduate Fellowship. X.-Q.\ Sun acknowledges support from the Gordon and Betty Moore Foundations EPiQS Initiative through Grant GBMF8691.
\end{acknowledgments}

\bibliography{DLA_quantum.bib}{}
\bibliographystyle{apsrev4-1}  
\end{document}


\title{Quantum entanglement and modulation enhancement of free-electron--bound-electron interaction -- Supplemental material}
\author{Zhexin Zhao$^1$}
\author{Xiao-Qi Sun$^{2,3}$}
\author{Shanhui Fan$^1$}
\affiliation{$^1$Department of Electrical Engineering, Ginzton Laboratory, Stanford University, Stanford, CA 94305, USA}
\affiliation{$^2$Department of Physics, McCullough Building, Stanford University, Stanford, CA 94305, USA}
\affiliation{$^3$Department of Physics, Institute for Condensed Matter Theory,
University of Illinois at Urbana-Champaign, IL 61801, USA}
\date{\today}

\begin{abstract}
We present a detailed derivation and discussion of the quantum theory of a single electron interacting with the two-level atom. 
\ZX{The coupling coefficient for the Tin-Vacancy center is presented as an example.}
We thoroughly investigate \ZX{a typical modulated electron in PINEM}, where the electron wave function consists of a single Gaussian wave packet modulated by a laser. \ZX{The optimal conditions of using a PINEM modulated free electron to measure the atomic coherence is presented. Discussions about experimental realization suggest that the proposed effects can be observed with current technology.} Details of the entanglement between distant two-level atoms as induced by electron-atom interaction are also presented. Finally, we discuss the interaction between the two-level atom and a dilute electron beam and provide comments about the semi-classical results. 
\end{abstract}

\maketitle

\section{Formalism}
\label{sec:quantum_theory}
In this section, we discuss the electron-atom interaction (Fig. 1 of the main text) from the first principle of quantum mechanics. We also explain the approximations leading to the simplified form of the interaction Hamiltonian. We study the low-velocity (non-retarded) limit, and then discuss the relativistic modification. 

In the low-velocity limit, the electrons are described by Schrodinger equation and the interaction potential is the Coulomb potential. The general Hamiltonian is
\begin{equation}
    \label{eq:quantum_H_general}
    \begin{split}
    H = &\, \int d^3\boldsymbol{r} \Psi^\dagger(\boldsymbol{r}) \Big[-\frac{\hbar^2 \nabla^2}{2m} + V_\textrm{ion}(\boldsymbol{r})\Big] \Psi(\boldsymbol{r}) \\ & + \int d^3\boldsymbol{r} \int d^3\boldsymbol{r}'\Psi^\dagger(\boldsymbol{r})\Psi(\boldsymbol{r})\frac{e^2}{4\pi\epsilon_0 |\boldsymbol{r} - \boldsymbol{r}'|}\Psi^\dagger(\boldsymbol{r}')\Psi(\boldsymbol{r}'),
    \end{split}
\end{equation}
where $V_\textrm{ion}$ is the potential of the positive ion of the atom, and the field operator
\begin{equation}
    \label{eq:psi_decompose}
    \Psi(\boldsymbol{r})  = \sum_i c_i \phi_i(\boldsymbol{r}),
\end{equation}
where $c_i$ is the annihilation operator of the mode $i$.
The eigen functions ($\phi_i$) of the non-interacting Hamiltonian (first term in Eq. (\ref{eq:quantum_H_general})) can represent either free electron states or bound electron states. The free electron states are close to free propagating electron states when the kinetic energy of the free electron is much larger than the magnitude of $V_\textrm{ion}$. 

We analyze the interaction part of the Hamiltonian $V_I$ (the last term in Eq. (\ref{eq:quantum_H_general})):
\begin{align}
\begin{split}
    \label{eq:quantum_VI}
    V_I = \sum_{ijlm} & c_i^\dagger c_j c_l^\dagger c_m \int d^3\boldsymbol{r} \int d^3\boldsymbol{r}'\phi_i^*(\boldsymbol{r})\phi_j(\boldsymbol{r})\\ & \times \frac{e^2}{4\pi\epsilon_0 |\boldsymbol{r} - \boldsymbol{r}'|}\phi_l^*(\boldsymbol{r}')\phi_m(\boldsymbol{r}').
\end{split}
\end{align}
Since we are mainly interested in the interaction between the free electron and the bound electron, we require one pair of the eigen functions represents the free electron states and the other pair represents the bound electron states. Explicitly, we consider two cases of Eq. (\ref{eq:quantum_VI}): (1) $\phi_i^*$, $\phi_j$ are free electron states, and $\phi_l^*$, $\phi_m$ are bound electron states. \TB{This is the contribution from the Coulomb integral}. (2) $\phi_i^*$, $\phi_m$ are free electron states, and $\phi_l^*$, $\phi_j$ are bound electron states. \TB{This is the contribution from the exchange integral. Exchange integral is a measure of the overlap of the wave functions between the bound electron state and the free electron state. In our case, the overlap is small since the free electron state and the bound electron state has very different spatial sizes. Therefore we neglect the exchange integral and only consider the contribution from the Coulomb integral.}  

We assume that the bound electron wave function is tightly confined in a space around the ion at $\boldsymbol{r}_a$. Thus, we can make the dipole approximation
\begin{equation}
    \label{eq:approximation_r}
    \frac{1}{|\boldsymbol{r} - \boldsymbol{r}'|} \approx \frac{1}{|\boldsymbol{r} - \boldsymbol{r}_a|} + \frac{(\boldsymbol{r} - \boldsymbol{r}_a) \cdot (\boldsymbol{r}' - \boldsymbol{r}_a)}{|\boldsymbol{r} - \boldsymbol{r}_a|^3}.
\end{equation}
With this approximation, the Hamiltonian in Eq. (\ref{eq:quantum_H_general}) takes the form:
\begin{align}
    \label{eq:quantum_Hamiltonian_2}
    \begin{split}
    H =&\, \sum_\alpha E_\alpha c_\alpha^\dagger c_\alpha + \sum_{k}E^\textrm{free}_k c_k^\dagger c_k + \sum_{kk'\alpha\alpha'} c_{k}^\dagger c_{k'} c_{\alpha}^\dagger c_{\alpha'}\\ & \times \int d^3\boldsymbol{r} \phi_k^*(\boldsymbol{r})\phi_{k'}(\boldsymbol{r})  \frac{e^2(\boldsymbol{r} - \boldsymbol{r}_a)\cdot \boldsymbol{l}_{\alpha\alpha'}}{4\pi\epsilon_0 |\boldsymbol{r} - \boldsymbol{r}_a|^3},
    \end{split}
\end{align}
where the effective dipole vector $\boldsymbol{l}_{\alpha\alpha'}$ is 
\begin{equation}
    \label{eq:effective_dipole_vec}
    \boldsymbol{l}_{\alpha\alpha'} = \int \phi_\alpha^*(\boldsymbol{r})(\boldsymbol{r} - \boldsymbol{r}_a)\phi_{\alpha'}(\boldsymbol{r})d\boldsymbol{r},
\end{equation}
which implies that $\boldsymbol{l}_{\alpha\alpha} = 0$ for any wave functions with inversion symmetry. For simplicity, we consider a two-level atom with two bound states $\phi_1$ and $\phi_2$. The interaction shown in Eq. (\ref{eq:quantum_Hamiltonian_2}) takes the form of $\boldsymbol{p}\cdot \boldsymbol{E}$, where $\boldsymbol{p}=e\boldsymbol{l}_{\alpha\alpha'}$ is the dipole moment of atomic transition from $\alpha'$ to $\alpha$, and $\boldsymbol{E}$ is the electric field generated by the free electron at the position of the atom. If relativistic corrections are considered, this electric field here, which is static, should be replaced by the retarded electric field.

\begin{figure}
    \centering
    \includegraphics[width=0.6\linewidth]{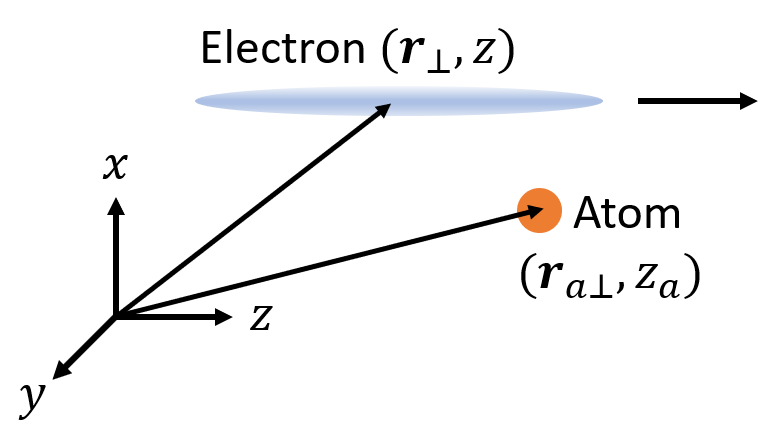}
    \caption{Schematic of a free electron passing near an atom. The free-electron wave function is $\phi(\boldsymbol{r})$, where $\boldsymbol{r} = (\boldsymbol{r}_\perp, z)$. The free electron propagates along the $z$-direction and localizes near $\boldsymbol{r}_{n_\perp}$ in the transverse dimension. The atom is at $\boldsymbol{r}_a = (\boldsymbol{r}_{a_\perp}, z_a)$}
    \label{fig:electron_atom}
\end{figure}

To describe the experiments where the free electron usually travels along a particular direction (denoted as the $z$-direction below), and is localized in the transverse directions (Fig. \ref{fig:electron_atom}), we choose the following basis for the free electron:
\begin{equation}
    \label{eq:phi_nkz}
    \phi_{n_\perp,k_z}(\boldsymbol{r}_\perp, z) = \frac{1}{\sqrt{L}} \phi_{n_\perp}(\boldsymbol{r}_\perp)e^{ik_z z},
\end{equation}
where $\phi_{n_\perp}(\boldsymbol{r}_\perp)$ describes the distribution of the wave function in the transverse coordinate $r_\perp$. We assume that $\phi_{n_\perp}$ is localized around a transverse position $\boldsymbol{r}_{n_\perp}$, $L$ is the length of the normalization box, which leads to the discretization of $k_z$. We assume that the two-level atom is at $\boldsymbol{r}_a = (\boldsymbol{r}_{a\perp}, z_a)$. The integration in the last term of Eq. (\ref{eq:quantum_Hamiltonian_2}) becomes:
\begin{equation}
    \label{eq:integration}
    \begin{split}
    & \frac{1}{L} \int dz e^{i(k_z' - k_z)z}\frac{e^2[\boldsymbol{r}_{n_\perp} - \boldsymbol{r}_{a\perp} + (z-z_a)\hat{e}_z]\cdot \boldsymbol{l}_{\alpha\alpha'}}{4\pi\epsilon_0[|\boldsymbol{r}_{n_\perp}-\boldsymbol{r}_{a\perp}|^2 + (z-z_a)^2]^\frac{3}{2}} \\ & = \frac{e^2(k_z' - k_z)}{2\pi\epsilon_0 L} \big\{-K_1\big[|(k_z'-k_z)(\boldsymbol{r}_{n_\perp}-\boldsymbol{r}_{a\perp})|\big] \hat{e}_{n\perp} + \\ & \;\;\;\; iK_0\big[|(k_z'-k_z)(\boldsymbol{r}_{n_\perp}-\boldsymbol{r}_{a\perp})|\big] \hat{e}_z \big\}\cdot \boldsymbol{l}_{\alpha\alpha'}e^{i(k_z' - k_z)z_a},
    \end{split}
\end{equation}
where $\hat{e}_{n\perp}$ is the unit vector parallel to $\boldsymbol{r}_{a\perp} -  \boldsymbol{r}_{n_\perp}$, and $K_0$, $K_1$ are the 0\textsuperscript{th} and 1\textsuperscript{st} order modified Bessel functions of the second kind. In obtaining Eq. (\ref{eq:integration}), we make the approximation that $\phi_{n_\perp}^*(\boldsymbol{r}_\perp) \phi_{n_\perp}(\boldsymbol{r}_\perp)\approx \delta(\boldsymbol{r}_\perp - \boldsymbol{r}_{n_\perp})$ and integrate over $\boldsymbol{r}_\perp$. Thus, the Hamiltonian becomes
\begin{align}
    \label{eq:quantum_Hamiltonian_3}
    \begin{split}
    H =&\, \sum_\alpha E_\alpha c_\alpha^\dagger c_\alpha + \sum_{k}E^\textrm{free}_k c_{k}^\dagger c_k + \sum_{n_\perp q} b_{n_\perp q} \\& \times \frac{e^2 q}{2\pi\epsilon_0 L}e^{iqz_a} \big\{K_1\big[|q(\boldsymbol{r}_{n_\perp}-\boldsymbol{r}_{a\perp})|\big] \hat{e}_{n\perp} \\ & +iK_0\big[|q(\boldsymbol{r}_{n_\perp}-\boldsymbol{r}_{a\perp})|\big] \hat{e}_z \big\}\cdot [ \boldsymbol{l}_{21}\sigma^+ + \boldsymbol{l}_{12}\sigma^-],
    \end{split}
\end{align}
where the electron ladder operator $b_{n_\perp q}$ is defined as: 
\begin{equation}
    \label{eq:electron_ladder_op}
    b_{n_\perp q} = \sum_{k_z}c_{n_\perp k_z}^\dagger c_{n_\perp (k_z+q)}^{ },
\end{equation}
where $c^\dagger_{n_\perp k_z}$ and $c_{n_\perp k_z}$ are the creation and annihilation operators of the free-electron state $\phi_{n_\perp, k_z}$.
The two-level atom operators $\sigma^+ = c_2^\dagger c_1^{ }$ and $\sigma^- = c_1^\dagger c_2^{ }$. Furthermore, when the transverse position of the electron beam center is fixed, for instance at $\boldsymbol{r}_{n_\perp} = (0, 0)$, we omit the subscript $n_\perp$ and introduce a complex  coupling coefficient with unit of energy:
\begin{equation}
    \label{eq:coupling_gij}
    \begin{split}
    g_{ij}(q) = &\; \frac{e^2 q}{2\pi\epsilon_0 L}e^{iqz_a}\big[- K_1(|q r_{a\perp}|)\hat{e}_{n\perp} \\ & + iK_0(|q r_{a\perp}|)\hat{e}_z\big]\cdot \boldsymbol{l}_{ij},
    \end{split}
\end{equation}
where $r_{a\perp} = |\boldsymbol{r}_{a\perp}|$. We also find that $g_{ij}(q) = g_{ji}^*(-q)$. Thus, we get the Hamiltonian to describe the interaction between a two-level atom and an electron centered at $\boldsymbol{r}_{a\perp}=(0,0)$ propagating along $z$-direction, which is Eq. (1) of the main text.
\begin{equation}
    \label{eq:quantum_Hamiltonian_final}
    H = H_0 + H_I,
\end{equation}
where
\begin{equation}
    \label{eq:quantum_Hamiltonian_final_0}
    \begin{split}
    H_0 = &\; \sum_\alpha E_\alpha c_\alpha^\dagger c_\alpha + \sum_{k}E^\textrm{free}_k c_k^\dagger c_k 
    \end{split}
\end{equation}
with $\alpha=1,2$ is the non-interacting part, and 
\begin{equation}
    \label{eq:quantum_Hamiltonian_final_I}
    H_I = \sum_{q} b_{q}[g_{21}(q)\sigma^+ + g_{12}(q)\sigma^-]
\end{equation}
is the interaction part. 

With the Hamiltonian, we proceed to consider the process of a free electron scattered by the two-level atom. The scattering matrix is formally expressed as
\begin{equation}
    \label{eq:Smatrix_formal}
    S = \mathcal{T}\exp\Big[-\frac{i}{\hbar} \int_{-\infty}^{\infty} dt H_I(t) \Big],
\end{equation}
where
\begin{equation}
    \label{eq:HI_interaction_picture}
    H_I(t) = e^{\frac{iH_0 t}{\hbar}}H_I e^{-\frac{iH_0 t}{\hbar}}
\end{equation}
is the interaction part of the Hamiltonian in the interaction picture.
The first order term in the scattering matrix is
\begin{align}
    \label{eq:Smatrix_first_order}
    \begin{split}
    S_{\alpha'k',\alpha k}^{(1)} = &\, -\frac{i}{\hbar}\int_{-\infty}^\infty dt e^{\frac{i(E_{\alpha'}+E_{k'}^\textrm{free})t}{\hbar}} \langle \alpha'k'|H_I|\alpha k\rangle \\ & \times e^{-\frac{i(E_{\alpha}+E_k^\textrm{free})t}{\hbar}} \\ = &\,-\frac{i}{\hbar}\langle \alpha'k'|H_I|\alpha k\rangle \\ & \times \delta\Big(\frac{E_{\alpha'}+E^\textrm{free}_{k'} - E_\alpha - E^\textrm{free}_k}{\hbar}\Big),
    \end{split}
\end{align}
where $\delta[(E_{\alpha'}+E^\textrm{free}_{k'} - E_\alpha - E^\textrm{free}_k)/\hbar]$ is a requirement of energy conservation.
Since we use a box length $L$ to discretize the electron momentum, i.e. $\Delta k_z = 2\pi/L$, and the local dispersion relation of an electron with central velocity $v_0$ is $\Delta \omega = v_0 \Delta k_z$, we can see that the $-\infty$ and $+\infty$ of the time corresponds to $-L/2v_0$ and $L/2v_0$ respectively, i.e. $\delta[(E_{\alpha'}+E^\textrm{free}_{k'} - E_\alpha - E^\textrm{free}_k)/\hbar]$ is replaced by $\delta_{(E_{\alpha'}+E^\textrm{free}_{k'}),(E_\alpha +E^\textrm{free}_k)}L/v_0$.
Therefore, 
\begin{equation}
    \label{eq:Smatrix_first_order_2}
    S^{(1)} = -\frac{iL}{\hbar v_0}\Big[ g_{21}\Big(\frac{\omega_a}{v_0}\Big) b_{\frac{\omega_a}{v_0}} \sigma^+ + g_{21}^*\Big(\frac{\omega_a}{v_0}\Big) b_{\frac{\omega_a}{v_0}}^\dagger \sigma^- \Big].
\end{equation}
We define a dimensionless coupling coefficient 
\begin{equation}
    \label{eq:coupling_g}
    \begin{split}
    g := &\; \frac{L}{\hbar v_0}g_{21}\Big(\frac{\omega_a}{v_0}\Big) \\ =&\; \frac{e^2 \omega_a}{2\pi\epsilon_0 \hbar v_0^2}e^{\frac{i\omega_a z_a}{v_0}}\Big[-K_1\Big(\Big|\frac{\omega_a r_{a\perp}}{v_0} \Big|\Big)\hat{e}_{n\perp} \\ & + iK_0\Big(\Big|\frac{\omega_a r_{a\perp}}{v_0} \Big|\Big)\hat{e}_z\Big]\cdot \boldsymbol{l}_{21}.
    \end{split}
\end{equation}
We comment that with the relativistic correction, the coupling coefficient should become
\begin{align}
    \label{eq:coupling_g_relativistic}
    \begin{split}
    g = &\; \frac{e^2\omega_a}{2\pi \epsilon_0 \gamma \hbar v_0^2}e^{\frac{i\omega_a z_a}{v_0}}\Big[-K_1\Big(\Big|\frac{\omega_a r_{a\perp}}{\gamma v_0} \Big|\Big)\hat{e}_{n\perp} \\ & + \frac{i}{\gamma}K_0\Big(\Big|\frac{\omega_a r_{a\perp}}{\gamma v_0} \Big|\Big)\hat{e}_z\Big]\cdot \boldsymbol{l}_{21},
    \end{split}
\end{align}
where $\gamma = 1/\sqrt{1 - v_0^2/c^2}$.
Furthermore, we drop the subscript $q =\frac{\omega_a}{v_0}$ of the electron ladder operator $b_q$. Thus, the first order perturbation of the scattering matrix is
\begin{equation}
    \label{eq:Smatrix_first_order_3}
    S^{(1)} = -i(g b \sigma^+ + g^* b^\dagger \sigma^-).
\end{equation}

The second order term of the scattering matrix is
\begin{align}
    \label{eq:Smatrix_second_order}
    \begin{split}
    S^{(2)}_{\alpha' k',\alpha k} = &\; -\frac{iL}{\hbar v_0} \delta_{(E_{\alpha'} + E^\textrm{free}_{k'}),(E_{\alpha} +  E^\textrm{free}_{k})}\times \\ & \sum_{\alpha_1 k_1}\frac{\langle \alpha' k' |H_I| \alpha_1 k_1\rangle \langle \alpha_1 k_1|H_I|\alpha k\rangle}{E_{\alpha'} + E^\textrm{free}_{k'} - E_{\alpha_1} - E^\textrm{free}_{k_1} + i\delta_E},
    \end{split}
\end{align}
where $\delta_E$ is an infinitesimal positive energy.
We find that if $\alpha' \neq \alpha$, the second order term would be zero. Thus, the only non-zero contribution is $\alpha' = \alpha$, and $k' = k$ which is due to energy conservation. Suppose $\alpha = 1$,
\begin{align}
    \label{eq:Smatrix_second_order_11}
    \begin{split}
    S^{(2)}_{1k, 1k} & = -\frac{iL}{\hbar^2 v_0}\sum_{k_1}\frac{g_{12}(k_1 - k)g_{21}(k - k_1)}{v_0(k - k_1) - \omega_a + i\delta_\omega} \\ & = -\frac{iL^2}{2\pi \hbar^2 v_0^2}\int dq\frac{g^*_{21}(q) g_{21}(q)}{q - \frac{\omega_a}{v_0} + i\delta_q} \\ & = -\frac{L^2|g_{21}(\frac{\omega_a}{v_0})|^2}{2\hbar^2 v_0^2} -\frac{iL^2}{2\pi \hbar^2 v_0^2} P.V. \int dq\frac{|g_{21}(q)|^2}{q - \frac{\omega_a}{v_0}},
    \end{split}
\end{align}
where P.V. stands for the principal value. When $g_{21}(q)$ changes slowly with $q$ near $q = \frac{\omega_a}{v_0}$, we anticipate that the second term in Eq. (\ref{eq:Smatrix_second_order_11}) to be small and neglect it in the following analysis. Similarly, we can get $S^{(2)}_{2k,2k}$. Therefore, the second order term of the scattering matrix is
\begin{equation}
    \label{eq:Smatrix_second_order_2}
    S^{(2)} \approx -\frac{1}{2}|g|^2 I,
\end{equation}
where $I$ is the identity operator. Thus, we get the scattering matrix up to second order:
\begin{equation}
    \label{eq:Smatrix_approximation}
    S\approx (1 - \frac{1}{2}|g|^2)I -i(gb\sigma^+ + g^* b^\dagger \sigma^-).
\end{equation}

\section{\label{sec:g_SnV}Coupling coefficient for the Tin-Vacancy center}

\begin{figure}
    \centering
    \includegraphics[width=0.96\linewidth]{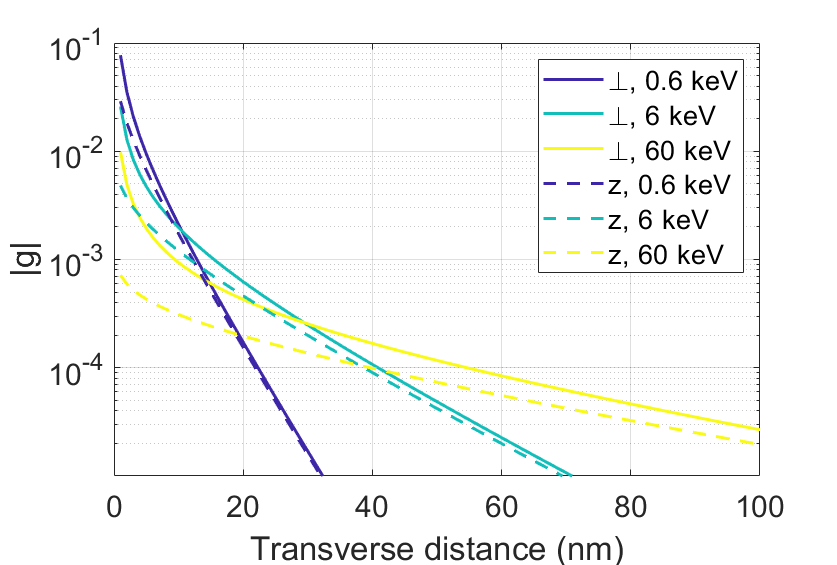}
    \caption{\ZX{The magnitude of the dimensionless coupling coefficient $g$ for the interaction between a SnV center and a free electron, as a function of the transverse separation distance. The solid and dashed curves represent the cases where the transition dipole is aligned perpendicular and parallel to the electron trajectory, respectively. The dark blue, green, and yellow curves are respectively for the electron with kinetic energy 0.6, 6, and 60 keV. }}
    \label{fig:g_SnV}
\end{figure}

\ZX{In this section, we investigate the coupling coefficient for the Tin-Vacancy (SnV) center interacting with free electrons. SnV has a nearly transform-limited emission bandwidth 35 MHz for the emission around 620 nm and its lifetime is 4.5 ns \cite{trusheim2020transform, rugar2020narrow}. The corresponding dipole length is 0.27 nm \cite{hilborn1982einstein}. The coupling coefficient $g$ is calculated using Eq. (\ref{eq:coupling_g_relativistic}).} 

\ZX{Figure \ref{fig:g_SnV} shows the coupling coefficient $g$ as a function of the transverse distance between the SnV and the free electron. The solid and dashed curves correspond to the cases where the transition dipole is aligned perpendicular and parallel to the electron trajectory, respectively. Each color represents a particular free-electron kinetic energy. Electrons with kinetic energy around 60 keV are widely accessible in transmission electron microscopes (TEMs) \cite{feist2015quantum}. For such electrons, the magnitude of the coupling coefficient is about $1\times10^{-3}$ when the transverse distance is about 10 nm for the perpendicularly oriented dipole. Furthermore, Fig. \ref{fig:g_SnV} shows that larger $|g|$ can be achieved with smaller transverse distance and smaller electron kinetic energy, which is consistent with \cite{yang2018maximal}.}

\section{\label{sec:guassian_electron} Interaction with a modulated Gaussian electron packet}

\begin{figure}
    \centering
    \includegraphics[width=0.98\linewidth]{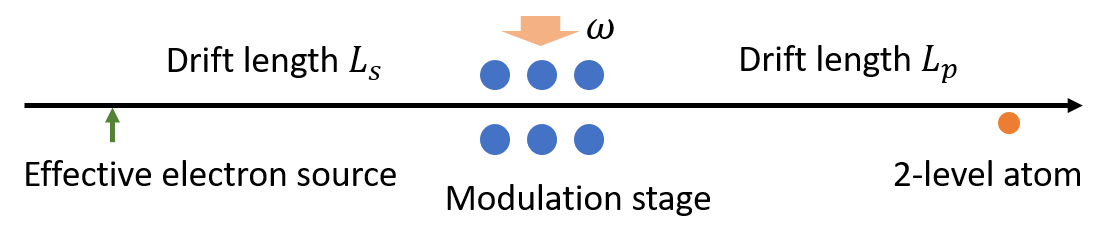}
    \caption{Schematic of a modulated electron wave packet interacting with the two-level atom. The effective electron source position is the Gaussian beam center in the longitudinal phase space, which is $L_s$ away from the energy modulation stage. The electron wave packet is modulated by a laser at frequency $\omega$ near a grating like structure. After the modulation, it drifts for a length $L_p$ before it interacts with the two-level atom.}
    \label{fig:guassian_wave_packet}
\end{figure}

In this section, we study an example of a single Gaussian electron wave packet, which is modulated by a laser and drifts for a length, interacting with a two-level atom, as shown in Fig. \ref{fig:guassian_wave_packet}. \ZX{This is the typical modulation that can be achieved in PINEM experiments \cite{feist2015quantum}.} 

The initial energy spread results from the energy uncertainty of the photo emission process. We choose the Gaussian envelope due to its simple mathematical expression, and for the fact that it occupies the minimal space in the longitudinal space-momentum phase space. We assume that, at the position of the electron source, the electron wave packet is unchirped, having the form:
\begin{equation}
    \label{eq:electron_gaussian_source}
    \Psi_e(q) = (2\pi\sigma_q^2)^{-\frac{1}{4}}\exp\Big(-\frac{q^2}{4\sigma_q^2}\Big)\exp(i\phi_0),
\end{equation}
where $\hbar q = p - p_0$ represents the deviation of the electron momentum from the central momentum $p_0 = \gamma m v_0$, ($\gamma = 1/\sqrt{1 - v_0^2/c^2}$), and $\sigma_q$ is the wave vector spread. \ZX{The initial phase $\phi_0\in[0, 2\pi]$ represents a random initial phase associated with the random electron emission time.}
The dispersion of the electron pulse can be determined from the dispersion relation of the free electron with central velocity $v_0$:
\begin{equation}
    \label{eq:electron_dispersion_relation}
    E(q) \approx \gamma m c^2 + \hbar v_0 q + \frac{\hbar^2q^2}{2\gamma^3 m}.
\end{equation}

We assume that the energy modulation stage is at a distance $L_s$ from the source. The electron wave function right before the modulation stage is 
\begin{equation}
    \label{eq:electron_gaussian_Ls}
     \Psi_e(q; L_s) = (2\pi\sigma_q^2)^{-\frac{1}{4}}\exp\Big[-\frac{q^2}{4\sigma_q^2}-i\frac{E(q)L_s}{\hbar v_0}+i\phi_0\Big].
\end{equation}

At the energy modulation stage, the near field of the some nano-structures, driven by a laser with frequency $\omega$, interacts with the electron and causes energy modulation \cite{feist2015quantum, black2019net, schonenberger2019generation}. Consistent with \cite{feist2015quantum}, we define the unit-less modulation strength as
\begin{equation}
    \label{eq:gm}
    g_m = \frac{e}{2\hbar \omega}\int dz E_{mz}(z)e^{-i\frac{\omega}{v_0}z},
\end{equation}
where $E_{mz}$ is the $z$-component of the modulation electric field.
We assume the length of the modulation stage is short such that the quadratic phase accumulated within the modulation stage, due to the free-electron dispersion, can be neglected. 
After the energy modulation,
\begin{align}
    \label{eq:electron_gaussian_modulation}
    \begin{split}
    & \Psi_e (q; L_s) = \;(2\pi\sigma_q^2)^{-\frac{1}{4}}\sum_n e^{in\phi_{g_m}+i\phi_0}J_n(2|g_m|) \times \\ & \exp\Big[-\frac{1}{4\sigma_q^2}\Big(q - n\frac{\omega}{v_0}\Big)^2 -iE\Big(q - n\frac{\omega}{v_0}\Big)\frac{L_s}{\hbar v_0}\Big],
    \end{split}
\end{align}
where $\phi_{g_m}$ is the phase of $g_m$, and $J_n$ is the $n^\textrm{th}$ Bessel function. During the energy modulation, the free electron interacts with the laser field and can absorb or emit integer number of photons. The process that the electron absorbs $n$ photons results in the n\textsuperscript{th} summand in the modulated free-electron wave function (Eq. (\ref{eq:electron_gaussian_modulation})). After the energy modulation, the electron propagates in the free space for a drift length $L_p$ (Fig. \ref{fig:guassian_wave_packet}). The wave function becomes
\begin{align}
    \label{eq:gaussian_electron_Lp}
    \begin{split}
    & \Psi_e (q; L_p) = \;(2\pi\sigma_q^2)^{-\frac{1}{4}}e^{-i\frac{E(q)L_p}{\hbar v_0}+i\phi_0}\sum_n e^{in\phi_{g_m}}J_n(2|g_m|)  \\ &\times \exp\Big[-\frac{1}{4\sigma_q^2}\Big(q - n\frac{\omega}{v_0}\Big)^2 -iE\Big(q - n\frac{\omega}{v_0}\Big)\frac{L_s}{\hbar v_0}\Big].
    \end{split} 
\end{align}

With this single electron wave function (Eq. (\ref{eq:gaussian_electron_Lp})) at the position of interaction with the two-level atom, we can obtain the perturbation on the two-level atom by studying $\langle b \rangle$ and $\langle b^2 \rangle$ for this incident electron state. \ZX{We can also get the explicit form of the EELS. To simplify the notation, we notice
\begin{align}
    \label{eq:E_difference}
    \begin{split}
    & E\Big(q+\frac{\omega_a}{v_0}-n\frac{\omega}{v_0}\Big) - E\Big(q-m\frac{\omega_a}{v_0}\Big) = \hbar v_0  \Big\{\Big [ \frac{\omega_a}{v_0} \\ & -(n-m)\frac{\omega}{v_0}\Big] +\zeta\Big[\Big(q+\frac{\omega_a}{v_0}-n\frac{\omega}{v_0}\Big)^2 - \Big(q-m\frac{\omega}{v_0}\Big)^2\Big]\Big\},
    \end{split}
\end{align}
where
\begin{equation}
    \label{eq:zeta}
    \zeta = \hbar /(2\gamma^3 m v_0).
\end{equation}
We firstly carry out the calculation of $s$:}
\begin{align}
    \label{eq:gaussian_electron_s}
    \begin{split}
    s = &\; \langle b \rangle = \int dq \Psi_e(q + \frac{\omega_a}{v_0})\Psi_e^*(q) \\ 
    = &\; (2\pi\sigma_q^2)^{-\frac{1}{2}} \sum_{nm}e^{i\phi_{g_m}(n - m)}J_n(2|g_m|)J_m(2|g_m|) \\ & \times \exp\Big\{-iL_p\frac{\omega_a}{v_0} - iL_s\Big[\frac{\omega_a}{v_0}-(n-m)\frac{\omega}{v_0}\Big] \Big\} \\ & \times \int dq  \exp\Big\{-\frac{1}{4\sigma_q^2} \Big[\Big(q + \frac{\omega_a}{v_0} - n\frac{\omega}{v_0}\Big)^2  \\ & \;\; + \Big(q - m\frac{\omega}{v_0}\Big)^2 \Big] -i\zeta L_s\Big[\Big(q + \frac{\omega_a}{v_0} - n\frac{\omega}{v_0}\Big)^2 \\ & \;\; - \Big(q - m\frac{\omega}{v_0}\Big)^2\Big] -i\zeta L_p \Big[\Big(q + \frac{\omega_a}{v_0}\Big)^2 - q^2\Big]\Big\}
    \end{split}
\end{align}    
\ZX{Integrate over $q$ and change the summation index to $l$ and $m$, where $n = m+l$, we get}
\begin{align}
    \label{eq:gaussian_electron_s_part2}
    \begin{split}
    s = &\; \sum_{lm}e^{il\phi_{g_m}}J_{m+l}(2|g_m|)J_m(2|g_m|) \exp\Big\{ \\ & -iL_p\frac{\omega_a}{v_0} - iL_s\Big(\frac{\omega_a}{v_0}-l\frac{\omega}{v_0}\Big) \\ & -\frac{1}{8\sigma_q^2}\Big(l\frac{\omega}{v_0} - \frac{\omega_a}{v_0}\Big)^2 -i\zeta L_p(2m+l)\frac{\omega}{v_0}\frac{\omega_a}{v_0} \\ & - 2\zeta^2\sigma_q^2\Big[L_s\Big(l\frac{\omega}{v_0} - \frac{\omega_a}{v_0}\Big) - L_p\frac{\omega_a}{v_0}\Big]^2\Big\}
    \end{split}
\end{align}
To sum over index $m$, we use the property of Bessel functions \cite{olver2010nist}:
\begin{align}
    \label{eq:bessel_function_1}
    J_\nu(u\pm v) & = \sum_{k=-\infty}^{+\infty}J_{\nu\mp k}(u)J_k(v), \\
    \label{eq:bessel_function_2}
    J_\nu(w)e^{i\nu \chi} &= \sum_{k=-\infty}^{\infty}J_{\nu+k}(u)J_k(v)e^{ik\alpha},
\end{align}
where $w$ is the third side of a triangle with two sides of length $u$ and $v$ and an angle $\alpha$ between the two sides, and angle $\chi$ is the angle facing the side of length $v$. The ideal drift length for perfect bunching given by classical analysis is 
\begin{equation}
    \label{eq:bunching_drift_length_classical}
    L_{p,c} = \frac{\gamma^3 m v_0^3}{2 |g_m|\hbar \omega^2} = \frac{v_0^2}{4\zeta |g_m|\omega^2}.
\end{equation}
With this notation (Eq. (\ref{eq:bunching_drift_length_classical})), and the properties of Bessel functions (Eqs. (\ref{eq:bessel_function_1}) and (\ref{eq:bessel_function_2})), we can simplify Eq. (\ref{eq:gaussian_electron_s}) as
\begin{align}
    \label{eq:gaussian_electron_s_2}
    \begin{split}
    s= &\;e^{-iL_p\frac{\omega_a}{v_0}} \sum_{l}e^{il(\phi_{g_m} + \frac{\pi}{2})} J_l\Big[4|g_m|\sin\Big( \frac{L_p}{4|g_m|L_{p,c}}\frac{\omega_a}{\omega} \Big)\Big] \\ & \times \exp\Big\{ -\frac{1}{8\sigma_q^2}\Big(\frac{\omega}{v_0}\Big)^2\Big(l - \frac{\omega_a}{\omega}\Big)^2+iL_s\frac{\omega}{v_0}\Big(l-\frac{\omega_a}{\omega}\Big) \\ &\;\; - 2\Big(\frac{\sigma_q}{4|g_m|}\frac{v_0}{\omega}\Big)^2\Big[\frac{L_s}{L_{p,c}}\Big(l - \frac{\omega_a}{\omega}\Big) - \frac{L_p}{L_{p,c}}\frac{\omega_a}{\omega}\Big]^2\Big\}.
    \end{split} 
\end{align}
Similarly, we can get $s_2$ that is the same expression as Eq. (\ref{eq:gaussian_electron_s_2}) but with $\omega_a$ replaced by $2\omega_a$. 

Equation (\ref{eq:gaussian_electron_s_2}) is general, including both on-resonant ($\omega_a =\omega$) and off-resonant excitation, as well as high-harmonic excitation ($\omega_a = l\omega, \,l > 1$). We find that the line width of the resonant excitation is determined by $\sigma_q$. To have a large contrast between on-resonant and off-resonant excitation, we want the electron to have small intrinsic energy spread. However, even with large energy spread, as long as the energy modulation $g_m$ is large enough, i.e. $|g_m| \gg \sigma_q v_0/\omega$, the magnitude of $s$ would not be reduced much compared with an electron with zero initial energy spread. In the case where the intrinsic energy spread approaches zero and the two-level atom transition frequency is on resonance with the $l^\textrm{th}$ harmonic of the modulation frequency, i.e. $\omega_a = l\omega$, we find that 
\begin{equation}
    \label{eq:s_on_resonance}
    s \approx  e^{-iL_p\frac{\omega_a}{v_0}+il(\phi_{g_m}+\frac{\pi}{2})} J_l\Big[4|g_\textrm{m}|\sin\Big(\frac{l L_p}{4|g_\textrm{m}|L_{p,c}}\Big)\Big].
\end{equation}

\ZX{We comment that the random initial phase of the wave function, which corresponds to the random emission time of the electron, has no influence on $s$. On the other hand, the phase of $s$ depends on the traveling time from the modulation stage to the two-level atom ($L_p/v_0$) (Eq. (\ref{eq:s_on_resonance})). The variance in this drift time would degrade the signal, if the detected signal requires the accumulation of multiple electrons, which we will discuss in Sec. \ref{sec:variance_propagation_time}. }

\ZX{The electron energy loss spectrum after such an electron interacting with a two-level atom in an arbitrary state can also be calculated using the free-electron wave function (Eq. (\ref{eq:gaussian_electron_Lp})) and Eq. (5) of the main text. 
\begin{equation}
    \label{eq:b_rhoe}
    \begin{split}
    & \langle k|b\rho_e^i | k\rangle = \Psi_e\big(k+\frac{\omega_a}{v_0}\big)\Psi_e^*(k) \\ & = (2\pi\sigma_q^2)^{-\frac{1}{2}} \sum_{nm}e^{i\phi_{g_m}(n - m)}J_n(2|g_m|)J_m(2|g_m|) \\ & \;\;\times \exp\Big\{-iL_p\frac{\omega_a}{v_0} - iL_s\Big[\frac{\omega_a}{v_0}-(n-m)\frac{\omega}{v_0}\Big] \Big\} \\ & \;\; \times  \exp\Big\{-\frac{1}{4\sigma_q^2} \Big[\Big(k + \frac{\omega_a}{v_0} - n\frac{\omega}{v_0}\Big)^2  \\ & \;\; + \Big(k - m\frac{\omega}{v_0}\Big)^2 \Big] -i\zeta L_s\Big[\Big(k + \frac{\omega_a}{v_0} - n\frac{\omega}{v_0}\Big)^2 \\ & \;\; - \Big(k - m\frac{\omega}{v_0}\Big)^2\Big] -i\zeta L_p \Big[\Big(k + \frac{\omega_a}{v_0}\Big)^2 - k^2\Big]\Big\}
    \end{split}
\end{equation}
To simplify the expression, we assume that the initial energy spread of the Gaussian wave packet is much smaller than the modulation energy ($\omega \gg v_0\sigma_q$) and the resonant condition $\omega_a = l\omega$ is satisfied, such that we can set $n = m+l$ in the summation. Thus, 
\begin{equation}
    \label{eq:b_rhoe_2}
    \begin{split}
    & \langle k|b\rho_e^i | k\rangle \approx (2\pi\sigma_q^2)^{-\frac{1}{2}} \exp\Big(-iL_p\frac{\omega_a}{v_0}+il\phi_{g_m}\Big) \times \\ & \sum_{m} J_{m+l}(2|g_m|)J_m(2|g_m|) \exp\Big\{ -\frac{1}{2\sigma_q^2} \Big(k - m\frac{\omega}{v_0}\Big)^2 \\& -i\zeta L_p \Big[2k\frac{\omega_a}{v_0} + \Big(\frac{\omega_a}{v_0}\Big)^2 \Big]\Big\}.
    \end{split}
\end{equation}
Similarly,
\begin{equation}
    \label{eq:b_rhoe_3}
    \begin{split}
    & \langle k|\rho_e^i b | k\rangle \approx (2\pi\sigma_q^2)^{-\frac{1}{2}} \exp\Big(-iL_p\frac{\omega_a}{v_0}+il\phi_{g_m}\Big) \times \\ & \sum_{m} J_{m}(2|g_m|)J_{m-l}(2|g_m|) \exp\Big\{ -\frac{1}{2\sigma_q^2} \Big(k - m\frac{\omega}{v_0}\Big)^2 \\& -i\zeta L_p \Big[2k\frac{\omega_a}{v_0} - \Big(\frac{\omega_a}{v_0}\Big)^2 \Big]\Big\}.
    \end{split}
\end{equation}
The EELS for arbitrary state of the two-level system is:
\begin{equation}
    \label{eq:EELS_e}
    \begin{split}
    & \Delta \rho_{e}(k) = -|g|^2 \rho_{e}^i(k) + |g|^2 \rho_{a11}\rho_{e}^i\big(k + \frac{\omega_a}{v_0}\big)\\ & + |g|^2 \rho_{a22}\rho_{e}^i\big(k-\frac{\omega_a}{v_0}\big)-i(2\pi\sigma_q^2)^{-\frac{1}{2}} \sum_n e^{-\frac{1}{2\sigma_q^2}(k-n\frac{\omega}{v_0})^2} \\ & \times \Big\{g \rho_{a12} e^{-i(L_p\frac{\omega_a}{v_0} + 2k \zeta L_p\frac{\omega_a}{v_0} -l\phi_{g_m})} \\ & \times \Big[J_{n+l}(2|g_m|) J_{n}(2|g_m|) e^{-i\zeta L_p(\frac{\omega_a}{v_0})^2} \\ & - J_n(2|g_m|)J_{n-l}(2|g_m|)e^{i\zeta L_p(\frac{\omega_a}{v_0})^2}\Big] - c.c. \Big\},
    \end{split}
\end{equation}
where c.c. stands for complex conjugate.
This expression provides a guidance to find the optimal conditions in a typical PINEM experiment to observe the spectral change that is to the first order of the coupling coefficient and the off-diagonal element of the atomic density matrix.}

\section{Conditions to optimize EELS signal for atomic coherence measurement}
\ZX{In this section, we continue the investigation of the modulated Gaussian electron wave packet in a typical PINEM experiment and search for the experimental conditions to optimize the EELS signal that can be used to measure the coherence of the two-level atom.}

\ZX{From Eq. (\ref{eq:EELS_e}), we find that, on the electron ladder, the EELS signal proportional to the first order of $g$ is
\begin{equation}
    \label{eq:EELS_e_g1}
    \Delta\rho_e^{(1)}\big(k=n\frac{\omega}{v_0}\big) = -i(2\pi\sigma_q^2)^{-\frac{1}{2}}[f_n(L_p) - f_n^*(L_p)],
\end{equation}
where
\begin{align}
    \label{eq:fn_k_Lp}
    \begin{split}
    f_n(L_p) = &\; g\rho_{a12}e^{-i(L_p\frac{\omega_a}{v_0}-l\phi_{g_m})}\\ & \times e^{-i2n\frac{\omega}{v_0}\zeta L_p\frac{\omega_a}{v_0}}\big[J_{n+l}J_n e^{-i\zeta L_p(\frac{\omega_a}{v_0})^2} \\ & -J_nJ_{n-l}e^{i\zeta L_p(\frac{\omega_a}{v_0})^2}\big].
    \end{split}
\end{align}
We use $J_n$ as a shorthand notation for $J_n(2|g_m|)$.}

\ZX{Since the PINEM spectra are generally symmetric with respect to the zero energy loss peak, to resolve the spectral change due to the coherence of the interacting two-level system, the spectral change should ideally be anti-symmetric, such that $\Delta \rho_{ej}(k) - \Delta \rho_{ej}(-k)$ would be the detected signal to reveal $\rho_{a12}$. 
The anti-symmetric part of the spectral change is maximized if $f_{-n}(L_p) = [f_{n}(L_p)]^*$, based upon Eq. (\ref{eq:EELS_e_g1}).
We notice that the term on the second and third line in Eq. (\ref{eq:fn_k_Lp}) becomes its complex conjugate multiplying $(-1)^{l+1}$ when $n \rightarrow -n$. 
Thus, to satisfy the condition $f_{-n}(L_p) = [f_{n}(L_p)]^*$, the first line of Eq. (\ref{eq:fn_k_Lp}) should be real when $l$ is odd (for instance, when the atomic transition frequency matches the modulation frequency, $l=1$), while it should be pure imaginary when $l$ is even, i.e.
\begin{equation}
    \label{eq:EELS_condition_max_asymmetry_1}
    \angle g + \angle \rho_{a12} - L_p\frac{\omega_a}{v_0} + l\phi_{g_m} = m\pi + \frac{l-1}{2}\pi,
\end{equation}
where $\angle$ represents the phase and $m$ is an integer.
For atom coherence $\rho_{a12}$ with arbitrary phase, we can tune $\phi_{g_m}$, which is the phase of the modulation laser field, to achieve condition Eq. (\ref{eq:EELS_condition_max_asymmetry_1}) and characterize the phase of $\rho_{a12}$.}

\begin{figure}
    \centering
    \includegraphics[width=1\linewidth]{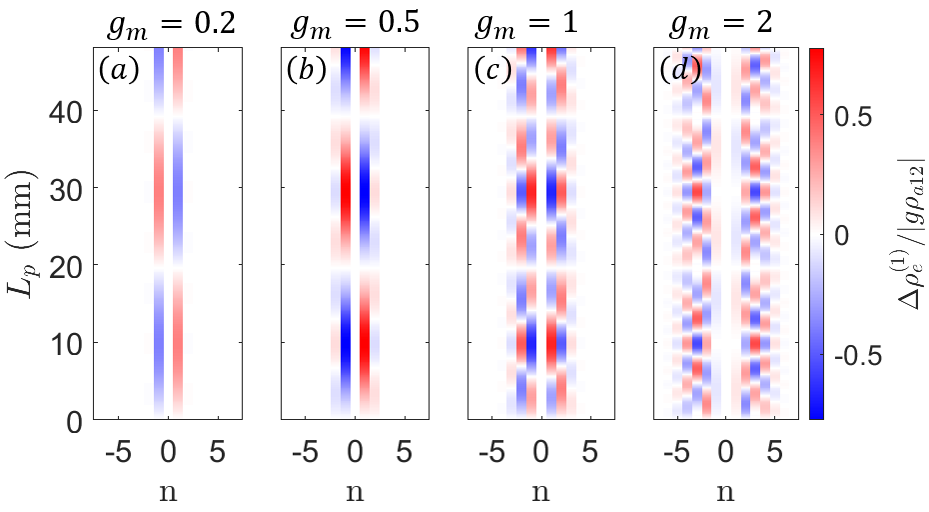}
    \caption{\ZX{The free-electron spectral change that is proportional to the first order of $g$ as a function of the drift length ($L_p$) for different modulation strengths: (a)-(d) are for modulation strength $g_m = 0.2, \, 0.5, \, 1, \, 2$ respectively. $n$ is the index of the electron ladder. The color represents $\Delta\rho_e^{(1)}/|g\rho_{a12}|$, which is the spectral integrated EELS for the $n$th peak that is proportional to the first order of coupling coefficient $g$, normalized to $|g\rho_{a12}|$. The free-electron initial kinetic energy is 60 keV and the interacting two-level system is SnV, as discussed in Sec. \ref{sec:g_SnV}. We also assume that condition Eq. (\ref{eq:EELS_condition_max_asymmetry_1}) is satisfied.}}
    \label{fig:EELS_Lp_sweep}
\end{figure} 

\ZX{Furthermore, we study the optimal modulation strength $g_m$ and the drift length from the modulation stage to the two-level system, such that the anti-symmetric part of the EELS is maximized. For simplicity, in the following discussion of this section, we consider the case where the modulation frequency and the transition frequency of the two-level system are the same, i.e. $\omega_a = \omega$ and $l=1$ in Eqs. (\ref{eq:EELS_e})-(\ref{eq:EELS_condition_max_asymmetry_1}). 
We find from Eq. (\ref{eq:fn_k_Lp}) that the EELS is a periodic function of the propagation length $L_p$. The periodicity is when
\begin{equation}
    \label{eq:Lp_periodic_condition}
    \zeta L_p\Big(\frac{\omega}{v_0}\Big)^2 = 2\pi.
\end{equation}
The corresponding propagation length is
\begin{equation}
    \label{eq:Lp_periodicity}
    L_p = \frac{2\pi v_0^2}{\zeta \omega^2} = \frac{4\pi\gamma^3 m v_0^3}{\hbar \omega^2}.
\end{equation}}

\ZX{Figure \ref{fig:EELS_Lp_sweep} shows the spectral change that is to the first order of $g$ ($\Delta\rho_e^{(1)}$) as a function of the propagation distance $L_p$ for a few different modulation strengths, where we assume that the condition to ensure the anti-symmetry of such spectral change (Eq. (\ref{eq:EELS_condition_max_asymmetry_1})) is satisfied. The periodic pattern associated with $L_p$ is observed for all modulation strengths and the periodicity has no dependence on the modulation strength. As a demonstration, we assume that the initial kinetic energy of the free electron is 60 keV and the two-level system is SnV with transition wavelength around 620 nm, as discussed in Sec. \ref{sec:g_SnV}. In such example, the periodicity of the propagation length is about 40 mm.  Usually, longer propagation length leads to larger uncertainty in propagation time and raises practical difficulties, we can choose $L_p \leq 2\pi\gamma^3 m v_0^3/\hbar \omega^2$. Furthermore, we find that the anti-symmetric part of the EELS is maximized when the drift length is $1/4$ of the periodicity, i.e. 
\begin{equation}
    \label{eq:EELS_Lp_optimal}
    L_{p,opt} =  \frac{\pi\gamma^3 m v_0^3}{\hbar \omega^2}.
\end{equation}}

\begin{figure}
    \centering
    \includegraphics[width=0.8\linewidth]{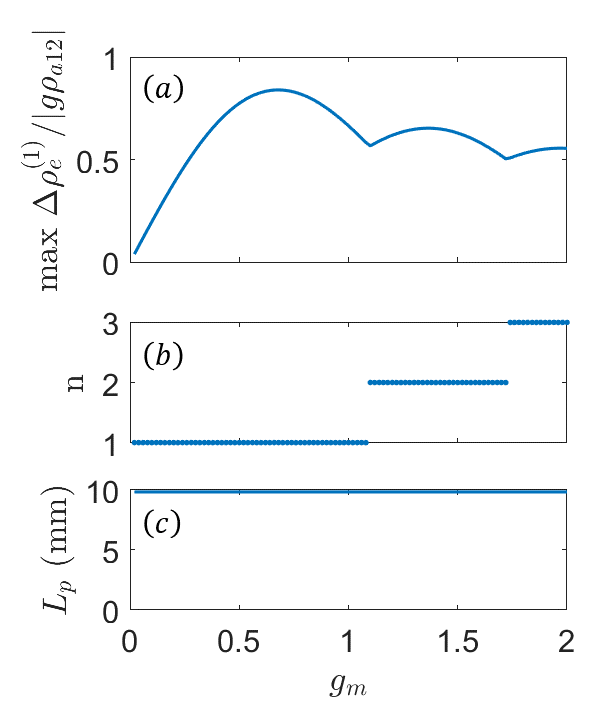}
    \caption{\ZX{(a) The maximal anti-symmetric part of the spectral change as a function of the modulation strength. The corresponding index of the electron ladder and the drift distance for the maximal $\Delta\rho_{e}^{(1)}$ are shown in (b) and (c), respectively. The optimal drift distance depends on the kinetic energy of the free electron and the properties of the two-level system, which are the same as those for Fig. \ref{fig:EELS_Lp_sweep}.}}
    \label{fig:EELS_gm_sweep}
\end{figure}

\ZX{Moreover, we numerically study the modulation strength that maximizes the anti-symmetric part of the spectral change. We assume condition Eq. (\ref{eq:EELS_condition_max_asymmetry_1}) is satisfied. Figure \ref{fig:EELS_gm_sweep}(a) shows the maximal spectral change as a function of the modulation strength. The corresponding index of the energy side-band of the free electron is shown in Fig. \ref{fig:EELS_gm_sweep}(b). When the modulation strength is small, the maximal anti-symmetric spectral change shows up in the first side-band. As the modulation strength increases, the maximal anti-symmetric spectral change appears in higher order side-bands. The propagation length to achieve the maximal anti-symmetric spectral change is shown in Fig. \ref{fig:EELS_gm_sweep}(c). It is independent of $g_m$ and agrees with Eq. \ref{eq:EELS_Lp_optimal} quantitatively. From Fig. \ref{fig:EELS_gm_sweep}(a), we find that the optimal modulation strength is $|g_m| = 0.68$. The corresponding PINEM spectrum and EELS are shown in Fig. 3 of the main text. }

\ZX{In summary, to optimize the EELS signal for measuring the atomic coherence, the phase of $g_m$ can be turned such that Eq. (\ref{eq:EELS_condition_max_asymmetry_1}) is satisfied. For $\omega_a = \omega$, the optimal modulation strength is $|g_m| = 0.68$ and the propagation distance from the modulation stage to the two-level system is given in Eq. (\ref{eq:EELS_Lp_optimal}).}

\section{Variance of the electron propagation time}
\label{sec:variance_propagation_time}
\ZX{In this section, we discuss the experimental feasibility for observing the predicted phenomena. As mentioned in Sec. \ref{sec:guassian_electron}, the random emission time of the free-electron has no influence on either the perturbation of the two-level system or the EELS, for such resonantly modulated free electrons. However, the fluctuation of the electron propagation time from the modulation stage to the two-level system can degrade the signal, if multiple electrons are involved for the detected signal. For instance, in EELS, the signal should be larger than the Poisson noise. The anti-symmetric part of the spectral change in EELS scales as $gN$, where $N$ is the total number of electrons, and the Poisson noise scales as $\sqrt{N}$. Therefore, the number of electrons $N>g^{-2}\approx 10^6$. Using lasers with MHz-GHz repetition rate, such number of electrons can be collected within a few seconds. Thus, we focus on the variance of the electron propagation time under the current experimental technology.}

\ZX{The propagation time would influence the phase of $s$ (Eq. (\ref{eq:s_on_resonance})), as well as spectral change $\Delta \rho_{e}(k)$ (Eq. (\ref{eq:EELS_e})), through the phase $L_{p}\omega_a/v_{0}$, where $L_{p}/v_{0}$ is the time for the free electron to propagate from the modulation stage to the two-level atom. The uncertainty of this propagation time comes mainly from the uncertainty in the initial velocity $\Delta v_0$ and the length of the trajectory $\Delta L_p$. Thus,
\begin{equation}
    \label{eq:phase_uncertainty}
    \begin{split}
    \Delta \phi & = \Delta \Big(\frac{\omega_a}{v_0}L_p \Big) \\ & \approx -L_p\frac{\omega_a}{v_0}\frac{\Delta E}{\beta \gamma^3 mc^2} + \frac{1}{2}(\Delta \theta)^2 L_p \frac{\omega_a}{v_0},
    \end{split}
\end{equation}
where $\Delta E$ is the initial energy spread of the free-electron, and $\Delta\theta$ is the angle divergence of the electron beam, which is related to the electron beam emittance.
To experimentally resolve the anti-symmetric electron energy spectral change, the condition $\Delta \phi \ll 2\pi$ should be satisfied. As an estimation using experimentally achievable parameters, the electron initial energy is $E_{kin} = 60$ keV ($\beta = 0.45$, $\gamma=1.12$) with $\Delta E = 0.5$ eV ($\Delta E/\beta \gamma^3 mc^2 = 1.6\times 10^{-6}$). We also assume that the electron beam has low emittance $\sim 0.1$ nm \cite{mcneur2016miniaturized} and divergence angle $\Delta \theta \sim 2$ mrad. The optimal drift length is $L_p \sim 10$ mm. For a atomic transition at wavelength 620 nm, the first term in Eq. (\ref{eq:phase_uncertainty}) is $0.057\times 2\pi$ and the second term in Eq. (\ref{eq:phase_uncertainty}) is $0.071\times 2\pi$, which are both smaller than $2\pi$. Thus, with the achievable experimental conditions, the predicted effects, including the coherent driving of the two-level atom and the anti-symmetric EELS change for detecting atomic coherence, would not be washed out.}

\section{Entanglement between 2 two-level systems}
\label{sec:entanglement_2atoms}
In this section, we show that two two-level atoms can have long range entanglement through the interaction with the same electron, as shown in Fig. 3 of the main text. We assume that the coupling coefficient between the free electron and the two-level atom 1(2) is $g_1$($g_2$). Suppose initially each atom is in a pure state. The initial state is 
\begin{equation}
    |\Psi^i\rangle = |\Psi_e^i\rangle |\Psi_{a1}^i\rangle |\Psi_{a2}^i\rangle .
\end{equation}
After the interaction with the two-level atom 1, the total state is,
\begin{equation}
\begin{split}
    |\Psi^{(1)}\rangle & = [S_1(g_1)\otimes I_2] |\Psi_e^i\rangle |\Psi_{a1}^i\rangle |\Psi_{a2}^i\rangle \\ & = \big[(1 - \frac{1}{2}|g_1|^2)|\Psi_e^i\rangle |\Psi_{a1}^i\rangle -ig_1b|\Psi_e^i\rangle \sigma^+|\Psi_{a1}^i\rangle \\ & \;\;\;\; -ig_1^*b^\dagger |\Psi_e^i\rangle \sigma^-|\Psi_{a1}^i\rangle \big]|\Psi_{a2}^i\rangle.
\end{split}
\end{equation}
After the interaction with the second two-level atom 2, the state becomes
\begin{equation}
\label{eq:entanglement_psi2}
\begin{split}
    & |\Psi^{(2)}\rangle =[S_{2}(g_2)\otimes I_{1}][S_{1}(g_1)\otimes I_{2}]|\Psi_e^i\rangle |\Psi_1^i\rangle |\Psi_2^i\rangle \\ & \approx |\Psi_e^i\rangle \big[(1-\frac{|g_1|^2 + |g_2|^2}{2})|\Psi_{a1}^i\rangle |\Psi_{a2}^i\rangle  \\ &\;\; - g_1g_2^*\sigma^+|\Psi_{a1}^i\rangle \sigma^-|\Psi_{a2}^i\rangle - g_1^*g_2\sigma^-|\Psi_{a1}^i\rangle \sigma^+|\Psi_{a2}^i\rangle \big] \\&\;\; -ib|\Psi_e^i\rangle \big[g_2|\Psi_{a1}^i\rangle \sigma^+|\Psi_{a2}^i\rangle + g_1 \sigma^+|\Psi_{a1}^i\rangle |\Psi_{a2}^i \rangle\big] \\&\;\;-i b^\dagger|\Psi_e^i\rangle \big[g_2^*|\Psi_{a1}^i\rangle \sigma^-|\Psi_{a2}^i\rangle + g_1^* \sigma^-|\Psi_{a1}^i\rangle |\Psi_{a2}^i \rangle\big]\\ &\;\; -g_1g_2 b^2|\Psi_e^i\rangle \sigma^+|\Psi_{a1}^i\rangle \sigma^+|\Psi_{a2}^i\rangle \\&\;\; -g_1^*g_2^*b^{\dagger 2}|\Psi_e^i\rangle \sigma^-|\Psi_{a1}^i\rangle \sigma^-|\Psi_{a2}^i\rangle,
\end{split}
\end{equation}
where we neglected terms with orders higher than $g^2$. When the two atoms are initially at the ground state and the free electron has an initial momentum $k^i$ with a small momentum spread, the final state is
\begin{equation}
\label{eq:entanglement_psi2_special}
\begin{split}
    & |\Psi^{(2)}\rangle \approx  (1-\frac{|g_1|^2 + |g_2|^2}{2})|k_e^i\rangle |1_{a1}\rangle |1_{a2}\rangle  \\ &\;\; -i\big|\big(k^i-\frac{\omega_a}{v_0}\big)_e\big\rangle \big(g_2|1_{a1}\rangle |2_{a2}\rangle + g_1 |2_{a1}\rangle |1_{a2} \rangle\big) \\ &\;\; -g_1g_2 \big|\big(k^i - 2\frac{\omega_a}{v_0}\big)_e \big\rangle |2_{a1}\rangle |2_{a2}\rangle.
\end{split}
\end{equation}
Thus, when the free electron is measured to have momentum $k^i - \omega_a/v_0$, the two-level atoms are in the entangled state
\begin{equation}
    \label{eq:entalgment_2atoms}
    |\Psi_{12}\rangle = \frac{1}{\sqrt{|g_1|^2 + |g_2|^2}}\big(g_2|1_{a1}\rangle |2_{a2}\rangle + g_1 |2_{a1}\rangle |1_{a2} \rangle\big) .
\end{equation}

\section{Interaction between the two-level system and a dilute electron beam}
\label{sec:supp_ebeam}

In this section, we provide detailed discussions about the problem that a dilute beam of modulated electrons interact with a two-level system in sequence. 
The adjacent electrons are separate enough such that, for each electron interacting with the two-level system, the influence of other electrons is negligible. Typically, this condition means that the separation is larger than the longitudinal size of the free-electron wave packet. When the coupling coefficient $g$ is small, complete control of the two-level system would require multiple free-electron scattering events. We assume that the time separation between adjacent electrons is $T$.

We focus on the dynamics of the two-level system. The effective dynamics of the two-level system is
\begin{equation}
    \label{eq:atom_effective_dynamics_supp}
    \frac{\Delta u}{\Delta t} = -i\frac{M}{T}u - \Gamma u, \; \Gamma= \begin{pmatrix} 0 & -\frac{1}{\tau} & 0 & 0 \\ 0 & \frac{1}{\tau} & 0 & 0 \\ 0 & 0 & \frac{1}{2\tau} & 0 \\ 0 & 0 & 0 & \frac{1}{2\tau} \end{pmatrix},
\end{equation}
where $u = [\rho_{a11},\, \rho_{a22},\, \rho_{a12},\, \rho_{a21}]^T$, matrix $M$ is given in Eq. (5) of the main text, and $\tau$ is the decay time of the two-level system. The effective Hamiltonian describing the two-level system driven by the electron beam is $H_\textrm{eff} = M/T-i\Gamma$. 
If the driving electron beam changes with time, the effective Hamiltonian becomes also time dependent. 

The steady state of the two-level system can be reached when the interaction time with the modulated electron beam is much longer than its decay time. The steady state corresponds to the eigen vector of the effective Hamiltonian associated with zero eigenvalue.
When the free electron is resonantly modulated, i.e. $\omega=\omega_a$, and drifts for a certain distance (Sec. \ref{sec:guassian_electron}), the two-level system is generically in a partially mixed state at steady state
\begin{equation}
    \label{eq:steady_state_supp}
    \rho_a = \frac{1}{1+2\big(\frac{2|gs|\tau}{T}\big)^2}\begin{bmatrix} 1+\Big(\frac{2|gs|\tau}{T}\Big)^2 & i\frac{2g^*s^*\tau}{T} \\ -i\frac{2gs\tau}{T} & \Big(\frac{2|gs|\tau}{T}\Big)^2 \end{bmatrix},
\end{equation}
where we neglect terms that are second order in $g$ in the effective Hamiltonian, i.e. we assume $|s|\gg|g|$. In the limit of fast decay $\tau\rightarrow 0$, the steady state approaches the ground state. In the opposite limit of slow decay $\tau\rightarrow \infty$, the steady state approaches a totally mixed state with equal probability for the ground and excited states, due to the de-coherence when tracing out the electron. The maximal off-diagonal element with magnitude $1/2\sqrt{2}$ is achieved when $2gs\tau/T = 1/\sqrt{2}$. Without modulation, i.e. $s = s_2= 0$, the steady state becomes $\rho_a = \frac{1}{1+2|g|^2\tau/T} \begin{bmatrix} 1+|g|^2\tau/T & 0 \\ 0 & |g|^2\tau/T \end{bmatrix}$, where the off-diagonal elements are zero and the excited state probability is typically much smaller in comparison with the modulated case.

\begin{figure}
    \centering
    \includegraphics[width=0.9\linewidth]{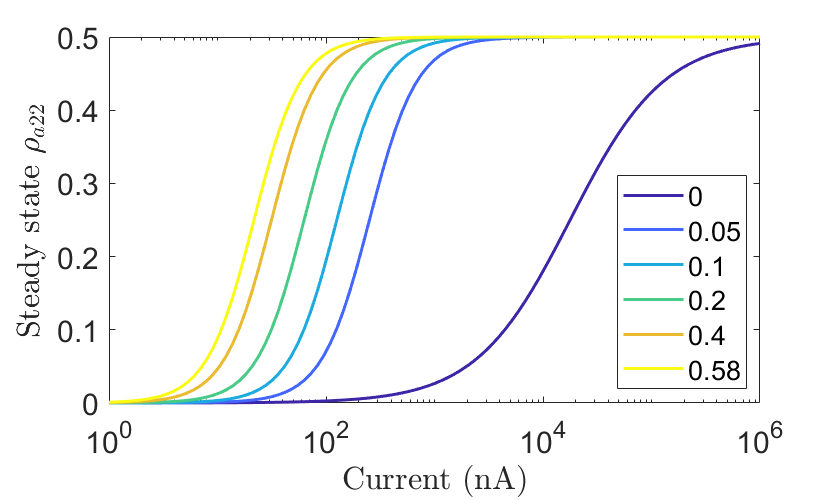}
    \caption{\ZX{The probability of the two-level system (SnV) in the steady state, as a function of the average current. The coupling coefficient $g=1\times10^{-3}$ and the decay time of SnV is $\tau=4.5$ ns. Curves with different colors represent different $\langle b\rangle$ of the free electon.}}
    \label{fig:steady_state}
\end{figure}

\ZX{We also numerically investigate the excited state probability in the steady state. This corresponds to a typical cathodoluminescence experiment, while the free electrons are modulated by a continuous-wave laser at the transition frequency before reaching the sample. We suppose the two-level system is SnV, as discussed in Sec. \ref{sec:g_SnV}, and the electron beam is focused within $\sim 10$ nm of SnV. The coupling coefficient is assumed to be $g=1\times10^{-3}$ and the decay time is $\tau=4.5$ ns. Figure \ref{fig:steady_state} shows the excited state probability in the steady state as a function of the average current, for different free-electron states characterized by $\langle b\rangle$. To saturate the excitation of the two-level system, the required average current with a modulated electron beam is about $g$ times the required current with an unmodulated electron beam. Below the saturation, the excited state probability under the excitation of a modulated free electron beam is much larger than that under the excitation of an unmodulated electron beam, with the same average current. Thus, we expect an increase of the cathodoluminescence signal with a modulated electron beam.}

Before reaching the steady state, the two-level system may experience Rabi oscillation. When the three nonzero eigenvalues of the effective Hamiltonian are all purely imaginary, the two-level system simply relaxes to the steady state. On the contrary, when two of the eigenvalues have nonzero real parts, the two-level system usually manifests oscillatory characteristics, which is the Rabi oscillation,  before reaching the steady state. The condition to enable such oscillation is
\begin{equation}
    \label{eq:rabi_oscillation_condition}
    \frac{T}{\tau}<8g|s|,
\end{equation}
which can be regarded as the strong coupling condition between a two-level system and an electron beam. This condition sets a minimal driving current threshold.

\ZX{To observe Rabi oscillation experimentally, the number of electrons interacting with the two-level system should surpass $1/8|gs|$ in a time duration much shorter than the lifetime of the two-level system. With a typical $|g|=10^{-3}$ and $|s|=0.58$, we find that the number of electrons should be larger than $2\times10^2$. The state-of-the-art nanotip electron emitters can emit thousands of electrons within several pico-seconds \cite{mcneur2016miniaturized, ceballos2019silicon}. Therefore, it is possible to observe the Rabi oscillation as a function of the number of emitted electrons per pulse experimentally. Nevertheless, the assumed experimental parameters are close to the state-of-the-art and might be challenging to achieve in practice. However, with the technology development, we believe the practical difficulties would soon be overcome.}

When the decay rate of the two-level system is small, i.e. $\frac{T}{\tau}\ll 8g|s|$, Rabi oscillation can be observed. The oscillation frequency is obtained from the eigenvalue of the effective Hamiltonian:
\begin{equation}
    \label{eq:Rabi_oscillation_freq}
    \Omega_{R} = \frac{2|gs|}{T},
\end{equation}
which is consistent with the semi-classical results  \cite{gover2020free}.

The consistency between our results and the semi-classical results can be explained as following. In the semi-classical approach, the electric field driving the two-level atom transition originates from the density distribution of the free electron. For resonant transition, the relevant quantity is the electric field at the resonant frequency, which is proportional to the time-varying current with the same frequency. Moreover, for electrons with velocity $v_0$, $t = z/v_0$, the resonant electric field is proportional to the Fourier component $k = \omega_a/v_0$ of the current density:
\begin{align}
    \label{eq:classical_quantum_connection}
    \begin{split}
    \int j(z)e^{-i\frac{\omega_a}{v_0}z}dz &= ev_0\int \langle \Psi^\dagger (z) \Psi(z) \rangle e^{-i\frac{\omega_a}{v_0}z}dz \\ & = ev_0 \Big\langle \frac{1}{2\pi} \int c_k^\dagger c_{k+\frac{\omega_a}{v_0}}^{} dk \Big \rangle \propto \langle b\rangle
    \end{split}
\end{align}
Therefore, we find that the classical interpretation of $s = \langle b\rangle$ is the amplitude of the current distribution with spatial frequency $k = \omega_a/v_0$. This connection provides a rigorous ground for the validity of semi-classical results related to Rabi oscillation.

\bibliography{DLA_quantum.bib}{}
\bibliographystyle{apsrev4-1}